\documentclass[aps,prd,preprint,showpacs,nofootinbib]{revtex4}
\usepackage{epsfig}
\begin{document}
\title{ Next-to-leading order QCD predictions for $A^0Z^0$
associated production at the CERN Large Hadron Collider}
\author{\center{Qiang Li, Chong Sheng Li\footnote{\hspace{-0.1cm}
Electronics address: csli@pku.edu.cn},
and Jian Jun Liu }} \affiliation{\small Department of Physics,
Peking University, Beijing 100871, China}
\author{Li Gang Jin}
 \affiliation{\small Institute of Theoretical Physics, Academia
Sinica, Beijing 100080, China}
\author{C.-P.Yuan}
\email{yuan@pa.msu.edu}
 \affiliation{\small
Department of Physics and Astronomy, Michigan State University,
East Lansing, MI 48824, USA}

\begin{abstract}
We present the calculations of the complete next-to-leading order
(NLO) QCD corrections (including supersymmetric QCD) to the
inclusive total cross sections of the associated production
processes $pp\rightarrow A^0Z^0+X$ in the Minimal Supersymmetric
Standard Model at the CERN Large Hadron Collider. Both the
dimensional regularization scheme and the dimensional reduction
scheme are used to organize the calculations which yield the same
NLO rates. The NLO correction can either enhance or reduce the
total cross sections, but it generally efficiently reduces the
dependence of the total cross sections on the
renormalization/factorization scale. We also examine the
uncertainty of the total cross sections due to the parton
distribution function uncertainties.
\end{abstract}


\maketitle
\section{Introduction}
The search for one or more Higgs bosons is the central task of the
CERN Large Hadron Collider (LHC), with $\sqrt{S}=14$ TeV and a
luminosity of 100 ${\rm fb^{-1}}$ per year. In the Standard Model
(SM), the Higgs boson mass is a free parameter with an upper bound
of $m_H\leq600$ --- 800\,GeV \cite{massh}. Beyond the SM, the
Minimal Supersymmetric Standard Model (MSSM), whose Higgs sector
is a special case of the Two Higgs Doublet Model (2HDM)
\cite{mssm}, is of particular theoretical interest, and contains
five physical Higgs bosons: two neutral CP-even bosons $h^0$ and
$H^0$, one neutral CP-odd boson $A^0$, and two charged bosons
$H^\pm$. The $h^0$ is the lightest, with a mass $m_{h^0}\leq140$\,
GeV when including the radiative corrections \cite{massh0}, and is
a SM-like Higgs boson especially in the decoupling region
($m_{A^0}\gg m_{Z^0}$). The other four are non-SM-like ones, and the
discovery of them may give the direct evidence of the MSSM. It has
been shown in \cite{detect,LHCHiggs} that the $h^0$ boson of MSSM
cannot escape detection at the CERN Large Hadron Collider (LHC)
 and that more than one neutral Higgs particle can be
found in large area of the supersymmetry (SUSY) parameter space

At the LHC, the neutral Higgs bosons can be produced through
following mechanisms: gluon fusion $gg \rightarrow \phi$
\cite{gg2h,gg2hnlo,gg2hnnlo,gg2hresum}, weak boson fusion $qq
\rightarrow qqV^*V^* \rightarrow qqh^0/qqH^0$ \cite{vv2h},
associated production with weak bosons \cite{wzh,v2vh,v22},
associated production with a heavy quark-antiquark pair
$gg,q\bar{q} \rightarrow t\bar{t}\phi/b\bar{b}\phi$ \cite{ttbbh}
and pairs production \cite{hpair}. Studying the associated
production process of a neutral Higgs boson and a vector boson at
future hadron colliders may be an interesting way in searching for
neutral Higgs bosons, since the total cross section may be large
and also the leptonic decay of the vector boson can be used as a
spectacular event trigger. In the SM, the process
$q\bar{q}^{(\prime )}\rightarrow W/Z^0 h^0_{SM}$ has been studied
both at the leading order (LO) \cite{wzh} and the
next-to-leading order (NLO) \cite{v22,wzhnlo}
in QCD. In the 2DHM and
 MSSM, the associated production of $h^0(H^0)Z^0$ and
$A^0Z^0$ has been studied only at tree level for Drell-Yan process
and at one-loop level for gluon fusion in \cite{hHZLO} and
\cite{AZLO,KaoAZLO,Kao}, respectively.

It was shown in Ref.~\cite{AZLO} that the $A^0Z^0$ associated
production rate at the LHC strongly depends on the SUSY parameters
$\tan{\beta}$ (the ratio of two vacuum expectation values) and
$m_A$ (the mass of $A^0$). The total cross section increases with
increment of $\tan{\beta}$, and decreases with increment of $m_A$.
In this paper, we present the complete NLO QCD, including
supersymmetric QCD, calculation for the cross section of the
associated production of $A^0Z^0$ through $b\bar{b}$ annihilation
process at the LHC. For simplicity, in our calculation, we neglect
the bottom quark mass except in the Yukawa couplings. Such
approximations are valid in all diagrams, in which the bottom
quark appears as an initial state parton, according to the
simplified Aivazis-Collins-Olness-Tung (ACOT) scheme \cite{acot}.
To regularize the ultraviolet (UV), soft and collinear
divergences, two regularization schemes are used in our
calculations for cross check, i.e. the dimensional regularization
(DREG) scheme \cite{DREG} (with naive $\gamma_5$ scheme
\cite{gamma5}) and the dimensional reduction (DRED) scheme
\cite{DRED}, and their results are compared.

The arrangement of this paper is as follows. In Sect.~II, we show
the LO results and define the notations. In Sect.~III, we present
the details of the calculations of both the virtual and real parts
of the NLO QCD corrections, and compare the results in DREG with
those in DRED. In Sect.~IV, by a detailed numerical analysis, we
present the predictions for the inclusive and differential cross
sections of the $A^0Z^0$ associated production
at the LHC. Sec.~V contains a brief conclusion.
For completeness, the relevant Feynman rules are collected in
Appendix~A, and the lengthy analytic expressions of the result
of our calculation are
summarized in Appendices B and C.

\section{Leading order calculations }
The related Feynman diagrams which contribute to the LO amplitude
of the partonic process $b(p_1) \bar{b}(p_2) \rightarrow
Z^0(p_3)A^0(p_4)$ are shown in Fig.~1. The LO amplitude in
$n=4-2\epsilon$ dimensions is
\begin{eqnarray}
&&M^{B}=\delta_{i_1i_2}\mu_r^{4-n}[M^{(s)}_0+M^{(t)}_0+M^{(u)}_0]\nonumber
\end{eqnarray}
with
\begin{eqnarray}
&&M^{(s)}_0=2im_b\bigg(\frac{A_1F_1}{s-m^2_{h^0}}+\frac{A_2F_2}{s-m^2_{H^0}}\bigg)\overline{v}(p_2)u(p_1)p_4.
\varepsilon_\rho(p_3) \, ,\nonumber\\
&&M^{(t)}_0=\frac{im_bA_3}{t}\overline{v}(p_2)\gamma^5(\not{\!p}_1-\not{\!p}_3)
\not{\!\varepsilon}(p_3)(C_V+C_A\gamma^5)u(p_1) \, ,\nonumber\\
&&M^{(u)}_0=\frac{im_bA_3}{u}\overline{v}(p_2)\not{\!\varepsilon}(p_3)(C_V+C_A\gamma^5)
(\not{\!p}_1-\not{\!p}_4)\gamma^5u(p_1),
\nonumber
\end{eqnarray}
where $\delta_{i_1i_2}$ is the color tensor ($i_1,i_2$ are color
indices for the initial state quarks), $\mu_r$ is a mass parameter
introduced to keep the couplings dimensionless,  $s,t$ and $u$ are
Mandelstam variables, which are defined as
\begin{eqnarray}
&& s=(p_1+p_2)^2, \qquad t=(p_1-p_3)^2, \qquad
u=(p_2-p_3)^2,\nonumber
\end{eqnarray}
 $A_{i=1,2,3}$, $F_{i=1,2}$ and $C_{V,A}$
 denote the coefficients appearing in the relevant
$h^0(H^0,A^0)b\bar{b}$, $h^0(H^0)Z^0A^0$ and $Z^0b\bar{b}$
couplings, respectively, and their explicit expressions are given
in Appendix A.

In order to simplify the expressions, we further introduce the
following Mandelstam variables:
\begin{eqnarray}
 &&  t'=t-m_{Z^0}^2,\ \ \ u'=u- m_{Z^0}^2.
\end{eqnarray}

After the $n$-dimensional phase space integration, the LO partonic
differential cross sections are given by
\begin{eqnarray}
&& \frac{d^2 \hat{\sigma}^B}{dt' du'} = \frac{\pi
S_{\epsilon}}{s^2\Gamma(1-\epsilon)} \bigg(\frac{t'u' - s
m_{Z^0}^2}{\mu_r^2s}\bigg)^{-\epsilon} \Theta(t'u' -s m_{Z^0}^2)
\Theta[s- (m_{Z^0} +m_{A^0})^2] \nonumber \\
&& \hspace{1.8cm} \times \delta (s+ t +u - m_{Z^0}^2 -m_{A^0}^2)
\overline{\sum}|M^B|^2
\end{eqnarray}
with
\begin{eqnarray}
&& \overline{\sum}{|M^B|}^2 =\frac{m^2_b}{6}
\Bigg\{\bigg[4m^2_{A^0}s-\frac{ (s-m_{Z^0}^2
-m_{A^0}^2)^2s}{m^2_{Z^0}}\bigg]\bigg(\frac{A_1F_1}{s-m^2_{h^0}}+\frac{A_2F_2}{s-m^2_{H^0}}\bigg)^2\nonumber
\\&&\hspace{2.0cm}
+A_3^2(|C_V|^2+|C_A|^2)\frac{2m^2_{Z^0}(1-\epsilon)(tu-m^2_{Z^0}m^2_{A^0})+st^2}{m^2_{Z^0}t^2}
\nonumber\\&&\hspace{2.0cm}+A_3^2(|C_V|^2+|C_A|^2)\frac{2m^2_{Z^0}(1-\epsilon)(tu-m^2_{Z^0}m^2_{A^0})+su^2}{m^2_{Z^0}u^2}\nonumber
\\&&\hspace{2.0cm}+\frac{4A_3C_As(t+u)(tu-m^2_{Z^0}m^2_{A^0})}{tum^2_{Z^0}}\bigg[\frac{A_1F_1}{(s-m^2_{h^0})}
+\frac{A_2F_2}{(s-m^2_{H^0})}\bigg]
\nonumber\\&&\hspace{2.0cm}+2A_3^2(|C_V|^2-|C_A|^2)\frac{2(1-\epsilon)m^2_{Z^0}(tu-m^2_{Z^0}m^2_{A^0})+s(2m^2_{Z^0}m^2_{A^0}-tu)}{m^2_{Z^0}tu}\Bigg\},
\end{eqnarray}
where $S_\epsilon=(4\pi)^{-2+\epsilon}$ and the $\Theta$ function
is the Heaviside step function. $\overline{\sum}{|M^B|}^2$ is the
LO squared matrix element of $b(x_1p_1) \bar{b}(x_2p_2)
\rightarrow Z^0(p_3)A^0(p_4)$, in which the colors and spins of
the outgoing particles have been summed, and the colors and spins
of the incoming ones have been averaged over.

The LO total cross section at the LHC is obtained by convoluting
the partonic cross section with the parton distribution
functions (PDFs) $G_{b,\bar{b}/p}$ in the proton:
\begin{eqnarray}
\sigma^B=\int dx_1dx_2
[G_{b/p}(x_1,\mu_f)G_{\bar{b}/p}(x_2,\mu_f)+ (x_1\leftrightarrow
x_2)]\hat{\sigma}^{B},\label{Born0}
\end{eqnarray}
where $\mu_f$ is the factorization scale and $\hat{\sigma}^{B}$ is
the Born level constituent cross section of $b(x_1p_1)
\bar{b}(x_2p_2) \rightarrow Z^0(p_3)A^0(p_4)$. Obviously, the
above LO results in the DREG scheme are equal to the ones in the
DRED scheme since the LO calculations are finite and free of any
singularity.

\section{Next-to-Leading order calculations}

The NLO contributions to the associated production of $A^0$ and
$Z^0$ can be separated into the virtual corrections arising from
loop diagrams of colored particles and the real corrections
arising from the radiation of a real gluon or a massless
(anti)quark. For both the virtual and real corrections, we will first
present the results in the DREG scheme, and then compare them with
the ones obtained in the DRED scheme.

\subsection{Virtual corrections}
The virtual corrections to $b\bar{b}\rightarrow A^0Z^0$ arise from
the Feynman diagrams shown in Fig.~2 and Fig.~3. They consist of
self-energy, vertex and box diagrams, which represent the SM QCD
corrections, arising from quarks and gluons, and supersymmetric QCD
corrections,
arising from squarks and gluinos. We carried out the calculation
in t'Hooft-Feynman gauge and used the dimensional regularization
in $n=4 -2\epsilon$ dimensions to regularize the ultraviolet (UV),
soft and collinear divergences in the virtual loop corrections. In
order to remove the UV divergences, we renormalize the
bottom quark masses in the Yukawa couplings and the wave function
of bottom quark, adopting the on-shell renormalization scheme
\cite{onmass}.

Denoting $m_{b0}$ and $\psi_{b0}$ as the bare bottom quark mass
and the bare wave function, respectively, the relevant
renormalization constants $\delta m_b, \delta Z_{bL}$ and $\delta
Z_{bR}$ are then defined as
\begin{eqnarray}
&& m_{b0}=m_b +\delta m_b,\\
 && \psi_{b0}=(1+\delta Z_{bL})^{\frac{1}{2}}\psi_{bL}+(1+\delta
 Z_{bR})^{\frac{1}{2}}\psi_{bR}.
\end{eqnarray}
After calculating the self-energy diagrams in Fig.~2, we obtain the
explicit expressions of all the renormalization constants as
follows:
\begin{eqnarray}
&& \frac{\delta m_b}{m_b} =  -\frac{\alpha_s}{4\pi} C_F
\bigg\{3B_0(m_b^2,0,m_b^2) - 2
\nonumber \\
&& \hspace{1.2cm} + \sum_{i=1}^2 \bigg[B_1
-\frac{m_{\tilde{g}}}{m_b} \sin2\theta_{\tilde{b}} (-1)^i
B_0\bigg](m_b^2,m_{\tilde{g}}^2,m_{\tilde{b}_i}^2)\bigg\},\nonumber
\\
&& \delta Z_{bL}= -\frac{\alpha_s}{2\pi}C_F\sum_{i=1}^2(R^{\tilde
b}_{i1})^2
(B_0+B_1)(0,m_{\tilde{b}_i}^2,m_{\tilde{g}}^2),\nonumber
\\
&& \delta Z_{bR}= -\frac{\alpha_s}{2\pi}C_F\sum_{i=1}^2(R^{\tilde
b}_{i2})^2
(B_0+B_1)(0,m_{\tilde{b}_i}^2,m_{\tilde{g}}^2),\nonumber
\end{eqnarray}
where $C_F=\frac{4}{3}$, $B_0,B_1$ are the scalar two-point
integrals \cite{denner}, $m_{\tilde{b}_{1,2}}$ are the sbottom
masses, $m_{\tilde{g}}$ is the gluino mass, and $R^{\tilde b}$ is
a $2\times 2$ matrix shown as below, which is defined to transform
the sbottom current eigenstates to the mass eigenstates
\cite{rotMa}:
\begin{equation}
\left(\begin{array}{c} \tilde{b}_1 \\ \tilde{b}_2 \end{array}
\right)= R^{\tilde{b}}\left(\begin{array}{c} \tilde{b}_L \\
\tilde{b}_R \end{array} \right), \ \ \ \ \
R^{\tilde{b}}=\left(\begin{array}{cc} \cos\theta_{\tilde{b}} &
\sin\theta_{\tilde{b}} \\ -\sin\theta_{\tilde{b}} &
\cos\theta_{\tilde{b}}
\end{array} \right)
\end{equation}
with $0 \leq \theta_{\tilde{b}} < \pi$, by convention.
Correspondingly, the mass eigenvalues $m_{\tilde{b}_1}$ and
$m_{\tilde{b}_2}$ (with $m_{\tilde{b}_1}\leq m_{\tilde{b}_2}$) are
given by
\begin{eqnarray}\label{Mq2}
\left(\begin{array}{cc} m_{\tilde{b}_1}^2 & 0 \\ 0 &
m_{\tilde{b}_2}^2 \end{array} \right)=R^{\tilde{b}}
M_{\tilde{b}}^2 (R^{\tilde{b}})^\dag, \ \ \ \ \
M_{\tilde{b}}^2=\left(\begin{array}{cc} m_{\tilde{b}_L}^2 & a_bm_b
\\ a_bm_b & m_{\tilde{b}_R}^2 \end{array} \right)
\end{eqnarray}
with
\begin{eqnarray}
m^2_{\tilde{b}_L} &=& M^2_{\tilde{Q}} +m_b^2
+m_Z^2\cos2\beta(I_{3L}^b -e_b\sin^2\theta_W), \\
m^2_{\tilde{b}_R} &=& M^2_{\tilde{D}} +m_b^2
+m_Z^2\cos2\beta e_b\sin^2\theta_W, \\
a_b &=& A_b -\mu\tan\beta.
\end{eqnarray}
Here, $M_{\tilde{b}}^2$ is the sbottom mass matrix.
$M_{\tilde{Q},\tilde{D}}$ and $A_{b}$ are soft
SUSY breaking
parameters and $\mu$ is the higgsino mass parameter . $I_{3L}^b$
and $e_b$ are the third component of the weak isospin (i.e. $-1/2$)
and the
electric charge of the bottom quark $b$ (i.e. $-1/3$), respectively.

The renormalized virtual amplitudes can be written as
\begin{eqnarray}
M^{V}=M^{unren}+M^{con}.
\end{eqnarray}
Here, $M^{unren}$ contains the radiative corrections
from the one-loop self-energy, vertex and box
diagrams, as shown in Fig.~2,
and $M^{con}$ is the corresponding counterterm.
Moreover, $M^{unren}$ can be separated into two parts:
\begin{eqnarray}
M^{unren}=\sum_{\alpha=a}^g M^\alpha + \sum_{\beta=a}^d M^{{\rm
box} (\beta)},
\end{eqnarray}
where $\alpha$ and $\beta$ denote the corresponding diagram
indexes in Fig.~2 and Fig.~3, respectively.
They can be further expressed as
\begin{eqnarray}
&& M^\alpha= \sum_{l=1}^{12} f_l^{\alpha} M_l,\label{fvertex}
\\
&& M^{box (\beta)}= \sum_{l=1}^{12} f_l^{box (\beta)}
M_l,\label{fbox}
\\
&& M^{unren}= \sum_{l=1}^{12} f_l^{unren} M_l, \hspace{1.0cm}
(f_l^{unren}=f_l^{\alpha}+f_l^{box (\beta)}) \, ,
\end{eqnarray}
where $f_l^\alpha$ and $f_l^{box(\beta)}$ are the form factors,
which are given explicitly in Appendix B, and the $M_l$ are the
standard matrix elements defined as
\begin{eqnarray}
&&M_{1(2)}=\bar{v}(p_2)u(p_1)p_{1(2)}.\epsilon(p_3),\nonumber
\\&&
M_{3(4)}=\bar{v}(p_2)P_{R}u(p_1)p_{1(2)}.\epsilon(p_3),\nonumber
\\&&
M_{5}=\bar{v}(p_2)\not{\!p}_3\not{\!
\epsilon}(p_3)u(p_1),\nonumber
\\&&
M_{6}=\bar{v}(p_2)\not{\!p}_3\not{\!
\epsilon}(p_3)P_Ru(p_1),\nonumber
\\&&
M_{7(8)}=\bar{v}(p_2)\not{\!p}_3P_Ru(p_1)p_{1(2)}.\epsilon(p_3),\nonumber
\\&&
M_{9}=\bar{v}(p_2)\not{\! \epsilon}(p_3)u(p_1),\nonumber
\\&&
M_{10}=\bar{v}(p_2)\not{\! \epsilon}(p_3)P_Ru(p_1),\nonumber
\\&&M_{11(12)}=\bar{v}(p_2)\not{\!p}_3u(p_1)p_{1(2)}.\epsilon(p_3).
\end{eqnarray}
The counterterm contribution
$M^{con}$ is separated into $M^{con(s)}$, $M^{con(t)}$ and
$M^{con(u)}$, i.e. the counterterms for s, t and u channels,
respectively, which are given by
\begin{eqnarray}
&& M^{con}= M^{con(s)} + M^{con(t)}+M^{con(u)},\nonumber
\\
&&
M^{con(s)}=2i\bigg(\frac{A_1F_1}{s-m^2_{h^0}}+\frac{A_2F_2}{s-m^2_{H^0}}\bigg)
\bigg[\delta m_b+\frac{m_b}{2}(\delta Z_{bL}+\delta
Z_{bR})\bigg](M_1+M_2) \, ,\nonumber\\
&& M^{con(t)}=\frac{-iA_3}{t} \bigg[\delta
m_b+\frac{m_b}{2}(\delta Z_{bL}+\delta
Z_{bR})\bigg][2(C_V-C_A)M_1-4C_VM_3\nonumber\\
&&\hspace{4.3cm}-(C_V-C_A)M_5+2C_VM_6] \, ,\nonumber
\\&& M^{con(u)}=\frac{iA_3}{u} \bigg[\delta
m_b+\frac{m_b}{2}(\delta Z_{bL}+\delta
Z_{bR})\bigg][2(C_V+C_A)M_2-4C_VM_4\nonumber\\
&&\hspace{4.3cm}+(C_V+C_A)M_5-2C_VM_6] \, .\nonumber
\end{eqnarray}

  The ${\cal O} (\alpha_s)$ virtual corrections to the differential
cross section can be expressed as
\begin{eqnarray}
&& \frac{d^2 \hat{\sigma}^V}{dt' du'} = \frac{\pi
S_{\epsilon}}{s^2\Gamma(1-\epsilon)}\bigg (\frac{t'u' - s
m_{Z^0}^2}{\mu_r^2s}\bigg)^{-\epsilon} \Theta(t'u' -s m_{Z^0}^2)
\Theta[s- (m_{Z^0} +m_{A^0})^2] \nonumber \\
&& \hspace{1.8cm} \times \delta (s+ t +u - m_{A^0}^2 -m_{Z^0}^2) \
2 \ {\rm Re} \bigg[\overline{\sum}{(M^V M^{B\ast})}\bigg],
\end{eqnarray}
where the renormalized amplitude $M^V$ is UV finite, but it still
contains the infrared (IR) divergences:
\begin{eqnarray}
M^V|_{IR} =\frac{\alpha_s}{2\pi}
\frac{\Gamma(1-\epsilon)}{\Gamma(1-2\epsilon)}
\bigg(\frac{4\pi\mu_r^2}{s}\bigg)^\epsilon
\bigg(\frac{A_2^V}{\epsilon^2} +\frac{A_1^V}{\epsilon}\bigg)M^B,
\end{eqnarray}
where \begin{eqnarray} A_2^V=-C_F,\qquad  A_1^V=-\frac{3}{2}C_F.
\end{eqnarray}
Here, the infrared divergences include the soft divergences and the
collinear divergences. The soft divergences are cancelled after
adding the real emission corrections, and the remaining collinear
divergences can be absorbed into the redefinition of PDF
\cite{altarelli}, which will be discussed in the following
subsections. Note that the coefficients $A_2^V$ and $A_1^V$ of the
infrared divergence terms are constants, similar to the Drell-Yan
type processes.
Needless to say that the SUSY QCD corrections do not
generate infrared divergences, for squarks and gluinos are massive
particles.

In the above calculation, we have adopted the naive $\gamma_5$
prescription in the DREG scheme to calculate the $A^0Z^0$
associated production rate. To cross check the above calculation,
we shall also adopt the DRED scheme to carefully treat the
$\gamma_5$ factor in the amplitude calculation. We shall show that
the total inclusive rate is independent of the regularization
scheme, though the individual contributions, from either virtual
or real emission corrections, can be scheme-dependent.

In the DRED scheme, $\delta Z_{bL}$
and $\delta Z_{bR}$ remain unchanged, however, $\delta m_b$ is
different, and
\begin{eqnarray}
 \bigg(\frac{\delta m_b}{m_b}\bigg)_{DREG}- \bigg(\frac{\delta m_b}{m_b}\bigg)_{DRED}
 = \frac{\alpha_s}{4\pi}
 C_F.\label{v1}
\end{eqnarray}
Similarly,
 the form factors are found to be different, and
\begin{eqnarray}
 &&{f_i^{unren}}_{DREG}-{f_i^{unren}}_{DRED}
 =-\frac{\alpha_s}{2\pi} C_F \,,
 \hspace{0.6cm} {\rm for} \quad i=1,2...6,\\
 &&{f_i^{unren}}_{DREG}-{f_i^{unren}}_{DRED} = 0  \,,
 \hspace{0.6cm} {\rm for} \quad i=7,8...12.
\end{eqnarray}
Thus, it is easy to obtain the following relations from the above
results:
\begin{eqnarray}
&&{M^V}_{DREG}-{M^V}_{DRED} = -\frac{\alpha_s}{4\pi}
 C_FM^B,\\
&&{\sigma^V}_{DREG}-{\sigma^V}_{DRED} = -\frac{\alpha_s}{2\pi}
 C_F\sigma^B+{\cal O}(\alpha^2_s).\label{0}
\end{eqnarray}

\subsection{Real gluon emission}

The Feynman diagrams for the real gluon emission process
$b(p_1)\bar{b}(p_2)\rightarrow Z^0(p_3)A^0(p_4)+g(p_5)$ are shown
in Fig.~4.

 The phase space integration for the real gluon emission
will produce infrared singularities, which can be either soft or
collinear and can be conveniently isolated by slicing the phase
space into different regions defined by suitable cut-offs. In this
paper, we use the two-cutoff phase space slicing method
\cite{cutoff} which introduces two small cut-offs to decompose the
three-body phase space into three regions.

First, the phase space can be separated into two regions by an
arbitrary small soft cut-off $\delta_s$, according to whether the
energy  ($E_5$) of the emitted gluon is soft, i.e. $E_5\leq
\delta_s\sqrt{s}/2$, or hard, i.e. $E_5> \delta_s\sqrt{s}/2$.
Correspondingly, the partonic real cross section can be written as
\begin{eqnarray}
\hat{\sigma}^{R}= \hat{\sigma}^{S} +\hat{\sigma}^{H},
\end{eqnarray}
where $\hat{\sigma}^{S}$ and $\hat{\sigma}^{H}$ are the
contributions from the soft and hard regions, respectively.
$\hat{\sigma}^{S}$ contains all the soft divergences, which can be
explicitly obtained after analytically
integrating over the phase space of
the emitted soft gluon. Second, in order to isolate the remaining
collinear divergences from $\hat{\sigma}^{H}$, we should introduce
another arbitrary small cut-off, called collinear cut-off
$\delta_c$, to further split the hard gluon phase space into two
regions, according to whether the Mandelstam variables satisfy the
collinear condition $-\delta_c s< u_{1,2}\equiv (p_{1,2}-p_5)^2<
0$ or not. Thus, we have
\begin{eqnarray}
\hat{\sigma}^{H}= \hat{\sigma}^{HC}+ \hat{\sigma}^{\overline{HC}},
\end{eqnarray}
where the hard collinear part $\hat{\sigma}^{HC}$ contains the
collinear divergences, which can be explicitly obtained after
analytically
integrating over the phase space of the emitted collinear gluon. The
hard non-collinear part $\hat{\sigma}^{\overline{HC}}$ is finite
and can be numerically computed using standard Monte-Carlo
integration techniques \cite{Monte}, and can be written in the
form:
\begin{eqnarray}
d\hat{\sigma}^{\overline{HC}}=\frac{1}{2s}
\overline{\sum}{|M^{b\bar{b}}|}^2 d\overline{\Gamma}_3.
\label{nonHC}
\end{eqnarray}
Here, $d\overline{\Gamma}_3$ is the hard non-collinear region of
the three-body phase space.

In the next two subsections, we will discuss in detail the soft
and hard collinear gluon emission.

\subsubsection{Soft gluon emission}

In the soft limit, i.e. when the energy of the emitted gluon is small,
with $E_5\leq \delta_s\sqrt{s}/2$, the matrix element squared
$\overline{\sum}{|M^{R}|}^2$
for the process
$b(p_1)\bar{b}(p_2)\rightarrow Z^0(p_3)A^0(p_4)g(p_5)$
can be simply factorized into the
Born matrix element squared times an eikonal factor $\Phi_{eik}$:
\begin{eqnarray}
\overline{\sum}{|M^{R}(b\bar{b}\rightarrow A^0Z^0 +g)|}^2
\stackrel{soft}{\rightarrow} (4\pi\alpha_s\mu_r^{2\epsilon})
\overline{\sum}{|M^{B}|}^2 \Phi_{eik},
\end{eqnarray}
where the eikonal factor $\Phi_{eik}$ is given by
\begin{eqnarray}
\Phi_{eik}= C_F\frac{s}{(p_1\cdot p_5)(p_2\cdot p_5)} .
\end{eqnarray}
Moreover, the phase space in the soft limit can also be factorized
as
\begin{eqnarray}
d\Gamma_3(b\bar{b}\rightarrow A^0Z^0 +g)
\stackrel{soft}{\rightarrow} d\Gamma_2(b\bar{b}\rightarrow A^0Z^0
) dS,
\end{eqnarray}
where $dS$ is the integration over the phase space of the soft
gluon, which is given by \cite{cutoff}
\begin{eqnarray}
dS =\frac{1}{2(2\pi)^{3- 2\epsilon}} \int_0^{\delta_s \sqrt{s}/2}
dE_5 E_5^{1 -2\epsilon} d \Omega_{2-2 \epsilon}.
\end{eqnarray}
Hence, the parton level cross section in the soft region can be
expressed as
\begin{eqnarray}\label{soft}
&&\hat{\sigma}^S =(4\pi\alpha_s\mu_r^{2\epsilon})\int
d\Gamma_2\overline{\sum}{|M^{B}|}^2 \int dS \Phi_{eik}.
\end{eqnarray}
Using the approach of Ref.~\cite{cutoff}, after analytically
 integrating
over the soft gluon phase space, Eq.~(\ref{soft}) becomes
\begin{eqnarray}
&&\hat{\sigma}^S =\hat{\sigma}^B \left [\frac{\alpha_s}{2\pi}
\frac{\Gamma(1-\epsilon)}{\Gamma(1-2\epsilon)}
\bigg(\frac{4\pi\mu_r^2}{s}\bigg)^\epsilon \right]
\bigg(\frac{A_2^s}{\epsilon^2} +\frac{A_1^s}{\epsilon}
+A_0^s\bigg)
\end{eqnarray}
with
\begin{eqnarray}
A_2^s=2C_F,\qquad A_1^s= -4C_F\ln\delta_s, \qquad
A_0^s=4C_F\ln^2\delta_s.
\end{eqnarray}

\subsubsection{Hard collinear gluon emission}
In the hard collinear region, i.e. $E_5> \delta_s\sqrt{s}/2$ and
$-\delta_c s< u_{1,2} < 0$, the emitted hard gluon is collinear to
one of the incoming partons. As a consequence of the factorization
theorems \cite{factor1}, the squared matrix element for
$b\bar{b}\rightarrow A^0Z^0 +g$ can be factorized into the product
of the Born squared matrix element and the Altarelli-Parisi
splitting function for $b(\bar{b})\rightarrow b(\bar{b})g$
 \cite{altarelli1,factor2}, i.e.
\begin{eqnarray}
\overline{\sum}{|M^{R}(b\bar{b}\rightarrow A^0Z^0 +g)|}^2
\stackrel{collinear}{\rightarrow} (4\pi\alpha_s \mu_r^{2\epsilon})
\overline{\sum}{|M^{B}|}^2 \bigg(\frac{-2P_{bb}(z,\epsilon)}{zu_1}
+\frac{-2P_{\bar{b}\bar{b}}(z,\epsilon)}{zu_2}\bigg),
\end{eqnarray}
where $z$ denotes the fraction of incoming parton $b(\bar{b})$'s
momentum carried by parton $b(\bar{b})$ with the emitted gluon
taking a fraction $(1-z)$, and $P_{ij}(z,\epsilon)$ are the
unregulated splitting functions in $n=4-2\epsilon$ dimensions for
$0<z<1$, which can be related to the usual Altarelli-Parisi
splitting kernels \cite{altarelli1} as
$P_{ij}(z,\epsilon)=P_{ij}(z) +\epsilon P_{ij}'(z)$. Explicitly
\begin{eqnarray}
&& P_{bb}(z)=P_{\bar{b}\bar{b}}(z)=C_F \frac{1
+z^2}{1-z}+C_F\frac{3}{2}\delta(1-z), \\
&&P_{bb}'(z)=P_{\bar{b}\bar{b}}'(z)= -C_F
(1-z)+C_F\frac{1}{2}\delta(1-z).\label{pp1}
\end{eqnarray}
Moreover, the three-body phase space can also be factorized in the
collinear limit, and, for example, in the limit $-\delta_c s< u_1
< 0$ it has the following form \cite{cutoff}:
\begin{eqnarray}
d\Gamma_3(b\bar{b}\rightarrow A^0Z^0 +g)
\stackrel{collinear}{\rightarrow} d\Gamma_2(b\bar{b}\rightarrow
A^0Z^0; s'=zs) \frac{(4\pi)^\epsilon}{16\pi^2\Gamma(1-\epsilon)}
dzdu_1[(z -1)u_1]^{-\epsilon}.
\end{eqnarray}
Here, the two-body phase space should be evaluated at the squared
parton-parton energy $zs$. Thus, the three-body cross section in
the hard collinear region is given by \cite{cutoff}
\begin{eqnarray}\label{eq:40}
&& d\sigma^{HC} =\hat{\sigma}^B \bigg[\frac{\alpha_s}{2\pi}
\frac{\Gamma(1-\epsilon)} {\Gamma(1-2\epsilon)}
(\frac{4\pi\mu_r^2}{s})^\epsilon\bigg] (-\frac{1}{\epsilon})
\delta_c^{-\epsilon}
\bigg[P_{bb}(z,\epsilon)G_{b/p}(x_1/z)G_{\bar{b}/p}(x_2) \nonumber
\\ && \hspace{1.4cm} + P_{\bar{b}\bar{b}}
(z,\epsilon)G_{\bar{b}/p}(x_1/z) G_{b/p}(x_2) +(x_1\leftrightarrow
x_2)\bigg] \frac{dz}{z} (\frac{1 -z}{z})^{-\epsilon} dx_1 dx_2,
\end{eqnarray}
where $G_{b(\bar{b})/p}(x)$ is the bare PDF.

\subsection{Massless (anti)quark emission}

In addition to the real gluon emission, a second set of real
emission corrections to the inclusive production rate of
$pp\rightarrow A^0Z^0$ at the NLO involves the processes with an
additional  massless (anti)quark in the final states:
\begin{eqnarray}
&& gb\rightarrow bA^0Z^0, \quad g\bar{b}\rightarrow
\bar{b}A^0Z^0 \, .\nonumber
\end{eqnarray}
The relevant Feynman diagrams for massless (anti)quark emission
(the diagrams for the antiquark emission are similar  and omitted
here) are shown in Fig.~5

Since the contributions from the real massless (anti)quark
emission contain the initial state collinear singularities, we
also need to use the two-cutoff phase space slicing method
\cite{cutoff} to isolate those collinear divergences.
Because there is no soft divergence in the splitting of
$g \to b \bar b$, we only need to separate the
phase space into two regions: the collinear region and the
hard non-collinear region.
 Thus, according to the approach shown in
Ref.~\cite{cutoff}, the cross sections for the processes with an
additional massless (anti)quark in the final states can be
expressed as
\begin{eqnarray}\label{accd}
&& d\sigma^{add}= \sum_{(\alpha=g,\beta=b,\bar{b})}
\hat{\sigma}^{\overline{C}}(\alpha\beta\rightarrow A^0Z^0 +X)
[G_{\alpha/p}(x_1) G_{\beta/p}(x_2) +(x_1\leftrightarrow x_2)]
dx_1dx_2 \nonumber
\\&& \hspace{1.0cm}
+\hat{\sigma}^B \bigg[\frac{\alpha_s}{2\pi}
\frac{\Gamma(1-\epsilon)} {\Gamma(1-2\epsilon)}
(\frac{4\pi\mu^2_r}{s})^\epsilon\bigg] (-\frac{1}{\epsilon})
\delta_c^{-\epsilon}
\bigg[P_{bg}(z,\epsilon)G_{g/p}(x_1/z)G_{\bar{b}/p}(x_2) \nonumber
\\ && \hspace{1.0cm} +G_{b/p}(x_1)P_{\bar{b}g}(z,\epsilon)G_{g/p}(x_2/z)+(x_1\leftrightarrow
x_2)\bigg] \frac{dz}{z} \bigg(\frac{1 -z}{z}\bigg)^{-\epsilon}
dx_1 dx_2,
\end{eqnarray}
where
\begin{eqnarray}
&& P_{bg}(z) =P_{\bar{b}g}(z)=\frac{1}{2}[z^2 +(1-z)^2],
\hspace{2.0cm} P_{bg}'(z)=P_{\bar{b}g}'(z)=-z(1-z).\label{pp2}
\end{eqnarray}
The first term in Eq.~(\ref{accd}) represents the non-collinear
cross sections for the two processes, which can be written in
the form:
\begin{eqnarray}
d\hat{\sigma}^{\overline{C}}=\frac{1}{2s}
\overline{\sum}{|M^{\alpha\beta}|}^2 d\overline{\Gamma}_3,
\label{qHC}
\end{eqnarray}
where $\alpha$ and $\beta$ denote the incoming partons in the
partonic processes, and $d\overline{\Gamma}_3$ is the three body
phase space in the non-collinear region. The second term in
Eq.~(\ref{accd}) represents the collinear singular cross sections.

\subsection{Mass factorization}
As mentioned above, after adding the renormalized virtual
corrections and the real corrections, the partonic cross sections
still contain the collinear divergences, which can be absorbed
into the redefinition of the PDF at NLO, in general called mass
factorization \cite{altarelli}. This procedure in practice means
that first we convolute the partonic cross section with the bare
PDF $G_{\alpha/p}(x)$, and then rewrite
$G_{\alpha/p}(x)$
in terms of the renormalized PDF
$G_{\alpha/p}(x,\mu_f)$in the numerical calculations. In the
$\overline{\rm MS}$ scheme, the scale dependent PDF
$G_{\alpha/p}(x,\mu_f)$ is given by \cite{cutoff}
\begin{eqnarray}
G_{\alpha/p}(x,\mu_f)= G_{\alpha/p}(x)+
\sum_{\beta}(-\frac{1}{\epsilon})\bigg [\frac{\alpha_s}{2\pi}
\frac{\Gamma(1 -\epsilon)}{\Gamma(1 -2\epsilon)} \bigg(\frac{4\pi
\mu_r^2}{\mu_f^2}\bigg)^\epsilon\bigg]  \int_x^1 \frac{dz}{z}
P_{\alpha\beta} (z) G_{\beta/p}(x/z).
\end{eqnarray}
After replacing the bare PDF by the renormalized $\overline{\rm MS}$
PDF and integrating out the collinear region of the phase space defined
in the two-cutoff phase space slicing method~\cite{cutoff}, the
resulting sum of Eq.~(\ref{eq:40}) and the collinear part (the second term) of
Eq.~(\ref{accd}) yields the remaining ${\cal O}$ collinear contribution
as~\cite{cutoff}:
\begin{eqnarray}
&& \sigma^{coll}= \int \hat{\sigma}^B\bigg[\frac{\alpha_s}{2\pi}
\frac{\Gamma(1-\epsilon)} {\Gamma(1-2\epsilon)}
\bigg(\frac{4\pi\mu^2_r}{s}\bigg)^\epsilon \bigg]
\{\tilde{G}_{b/p}(x_1,\mu_f) G_{\bar{b}/p}(x_2,\mu_f) +
G_{b/p}(x_1,\mu_f) \tilde{G}_{\bar{b}/p}(x_2,\mu_f) \nonumber
\\ && \hspace{1.2cm}
+\sum_{\alpha=b,\bar{b}}\bigg[\frac{A_1^{sc}(\alpha\rightarrow
\alpha g)}{\epsilon} +A_0^{sc}(\alpha\rightarrow \alpha
g)\bigg]G_{b/p}(x_1,\mu_f) G_{\bar{b}/p}(x_2,\mu_f) \nonumber
\\ && \hspace{1.2cm}
+(x_1\leftrightarrow x_2)\} dx_1dx_2,\label{11}
\end{eqnarray}
where
\begin{eqnarray}
&& A_1^{sc}(b\rightarrow bg)=A_1^{sc}(\bar{b}\rightarrow \bar{b}g)=C_F(2\ln\delta_s +3/2), \\
&& A_0^{sc}=A_1^{sc}\ln(\frac{s}{\mu_f^2}), \\
&&
\tilde{G}_{\alpha(=b,\bar{b})/p}(x,\mu_f)=\sum_{\beta=g,\alpha}\int_x^{1-
\delta_s\delta_{\alpha\beta}} \frac{dy}{y}
G_{\beta/p}(x/y,\mu_f)\tilde{P}_{\alpha\beta}(y)
\end{eqnarray}
with
\begin{eqnarray}
\tilde{P}_{\alpha\beta}(y)=P_{\alpha\beta}(y) \ln(\delta_c
\frac{1-y}{y} \frac{s}{\mu_f^2}) -P_{\alpha\beta}'(y).
\end{eqnarray}

The NLO total cross section for $pp\rightarrow A^0Z^0$ in
the $\overline{MS}$ factorization scheme is
obtained by summing up the
Born, virtual, soft, collinear and hard non-collinear
 contributions. In terms of the above notations, we have
\begin{eqnarray}
&& \sigma^{NLO}= \int dx_1dx_2 \{
\bigg[G_{b/p}(x_1,\mu_f)G_{\bar{b}/p}(x_2,\mu_f)+
(x_1\leftrightarrow x_2)\bigg](\hat{\sigma}^{B} +
\hat{\sigma}^{V}+ \hat{\sigma}^{S} +\hat{\sigma}^{\overline{HC}})\}
+\sigma^{coll} \nonumber
\\ && \hspace{0.4cm} +\sum_{(\alpha=g,\beta=b,\bar{b})}\int dx_1dx_2
\bigg[G_{\alpha/p}(x_1,\mu_f) G_{\beta/p}(x_2,\mu_f)
+(x_1\leftrightarrow x_2)\bigg]
\hat{\sigma}^{\overline{C}}(\alpha\beta\rightarrow A^0Z^0
+X) \, .\label{t}
\end{eqnarray}
We note that the above expression contains no singularities, for
$2A_2^V +A_2^s =0$ and $2A_1^V +A_1^s +A_1^{sc}(b\rightarrow bg)
+A_1^{sc}(\bar{b}\rightarrow \bar{b}g) =0$.
Namely, all the $1/{\epsilon^2}$ and $1/{\epsilon}$ terms cancel in
$\sigma^{NLO}$. The apparent logarithmic $\delta_s$ and $\delta_c$ dependent terms
also cancel with the
the hard non-collinear cross section $\hat{\sigma}^{\overline{HC}}$
after numerically integrating over its relevant phase space volume.

\subsection{Real emission corrections and NLO total cross sections in the DRED scheme}

In the end of Sec.~III A, cf. Eqs.~(\ref{v1})--(\ref{0}), we
discussed the results of virtual corrections in the DRED scheme.
Here, we examine the real emission corrections and the NLO total
cross section in the DRED scheme and compare them with those
obtained in the DREG scheme. We find that the contributions from
soft gluon emission remain the same, while the ones from hard
collinear gluon emission and massless (anti)quark emission are
different due to the difference in the parton
splitting functions and the perturbative PDFs.

First, the splitting functions in the DRED scheme contain no
$\epsilon$ parts, so that
\begin{eqnarray}
P_{ij}(z,\epsilon)_{DRED}=P_{ij}(z).\label{12}
\end{eqnarray}
Thus, from Eqs.~(\ref{11}) and (\ref{12}), we find the difference
\begin{eqnarray}
&&{\sigma^{coll}}_{DREG} - {\sigma^{coll}}_{DRED} =
-\frac{\alpha_s}{2\pi} \int \hat{\sigma}^B
\{\sum_{\beta}\int_{x_1}^{1-\delta_s\delta_{b\beta}} \frac{dy}{y}
G_{\beta/p}(x_1/y,\mu_f){P^\prime}_{b\beta}(y)
G_{\bar{b}/p}(x_2,\mu_f)\nonumber
\\&& \hspace{0.2cm}+ \sum_{\beta}\int_{x_2}^{1-\delta_s\delta_{\bar{b}\beta}} \frac{dy}{y}
G_{\beta/p}(x_2/y,\mu_f){P^\prime}_{\bar{b}\beta}(y)
G_{b/p}(x_1,\mu_f) +(x_1\leftrightarrow x_2)\}
dx_1dx_2 \, . \label{1}
\end{eqnarray}
Secondly, the perturbative PDFs defined in the DRED and DREG schemes are
different, and ~\cite{diffPDF}:
\begin{eqnarray}
G_{\alpha/p}(x,\mu_f)_{DREG}-G_{\alpha/p}(x,\mu_f)_{DRED}
=\frac{\alpha_s}{2\pi}\sum_{\beta}
\int_{x}^{1}\frac{dy}{y}{P^\prime}_{\alpha\beta}(x/y)G_{\alpha/p}(x,\mu_f)_{DREG}.
\end{eqnarray}
After substituting them into the formula for calculating the Born level
cross sections, cf. Eq.~(\ref{Born0}), we find the difference
arising from the perturbative PDFs, at the ${\cal O}(\alpha_s)$
level, as:
\begin{eqnarray}
&&{\sigma^{B}}_{DREG}-{\sigma^{B}}_{DRED}=
\frac{\alpha_s}{2\pi} \int \hat{\sigma}^B
\{\sum_{\beta}\int_{x_1}^{1} \frac{dy}{y}
G_{\beta/p}(x_1/y,\mu_f)_{DRED}{P^\prime}_{b\beta}(y)
G_{\bar{b}/p}(x_2,\mu_f)_{DRED}\nonumber
\\&& \hspace{0.2cm}+ \sum_{\beta}\int_{x_2}^{1} \frac{dy}{y}
G_{\beta/p}(x_2/y,\mu_f)_{DRED}{P^\prime}_{\bar{b}\beta}(y)
G_{b/p}(x_1,\mu_f)_{DRED} +(x_1\leftrightarrow
x_2)\} dx_1dx_2.\label{2}
\end{eqnarray}
Except the upper limit of the integral over y, the two expressions in
Eqs.~(\ref{1}) and (\ref{2}) are the same. After substituting
Eqs.~(\ref{1}), (\ref{2}) and (\ref{0}) into Eq.~(\ref{t}), we
find the relation between the two NLO total cross sections,
separately calculated in the DREG and DRED schemes, as follows:
\begin{eqnarray}
&&{\sigma^{NLO}}_{DREG} - {\sigma^{NLO}}_{DRED} =
\frac{\alpha_s}{2\pi} \int \hat{\sigma}^B
\{\sum_{\beta}\int_{1-\delta_s\delta_{b\beta}}^{1} \frac{dy}{y}
G_{\beta/p}(x_1/y,\mu_f){P^\prime}_{b\beta}(y)
G_{\bar{b}/p}(x_2,\mu_f)\nonumber
\\&& \hspace{1.2cm}+ \sum_{\beta}\int_{1-\delta_s\delta_{\bar{b}\beta}}^{1} \frac{dy}{y}
G_{\beta/p}(x_2/y,\mu_f){P^\prime}_{\bar{b}\beta}(y)
G_{b/p}(x_1,\mu_f) +(x_1\leftrightarrow x_2)\}
dx_1dx_2\nonumber
\\&& \hspace{1.2cm}-\frac{\alpha_s}{2\pi} C_F \sigma^B_{DRED}+{\cal O}
(\alpha^2_s).
\end{eqnarray}
Using the explicit expressions of the $\epsilon$ parts of the
splitting functions $P^\prime$, cf. Eqs.~(\ref{pp1}) and
(\ref{pp2}), we find
\begin{eqnarray}
\sigma^{NLO}_{DREG}=\sigma^{NLO}_{DRED}+{\cal O} (\alpha^2_s).
\end{eqnarray}
As expected, both schemes yield the same
NLO total cross sections, up to ${\cal O} (\alpha^2_s)$.

\subsection{Differential cross sections in transverse momentum and invariant mass}

In this subsection ,we present the differential cross section in
the transverse momentum of $Z^0$ and $A^0$ bosons, respectively,
and the invariant mass of the $A^0Z^0$ pair.
 Using the
notations defined in Ref.~\cite{beenakker2}, the differential
distribution of the transverse momentum
($p_T$) and rapidity ($y$ ) of $Z^0$ boson for the
processes
\begin{eqnarray}
p(p_1)+ p(p_2) \rightarrow Z^0 (p_3) +A^0 (p_4) \/  [+
g(p_5)/b(p_5)/\bar{b}(p_5)]
\end{eqnarray}
is given by
\begin{eqnarray} \label{integralpt}
\frac{d^2\sigma}{dp_T dy} =2 p_T S \sum_{\alpha,\beta}
\int_{x_1^-}^1 dx_1 \int_{x_2^-}^1 dx_2 x_1
G_{\alpha/p}(x_1,\mu_f) x_2 G_{\beta/p}(x_2,\mu_f) \frac{d^2
\hat{\sigma}_{\alpha\beta}}{dt' du'},
\end{eqnarray}
where $\sqrt{S}$ is the total center-of-mass energy of the
collider, and
\begin{eqnarray}
&& p_T^2= \frac{T_2U_2}{S} -m_{Z^0}^2, \hspace{1.8cm}
y=\frac{1}{2}\ln(\frac{T_2}{U_2}), \nonumber \\ && x_1^-=
\frac{-T_2 -m_{Z^0}^2 +m_{A^0}^2}{S +U_2}, \ \ \ \ \ \ x_2^-=
\frac{-x_1U_2 -m_{Z^0}^2 +m_{A^0}^2}{x_1S +T_2}
\end{eqnarray}
with $T_2=(p_2-p_3)^2 - m_{Z^0}^2$ and $U_2=(p_1-p_3)^2 -
m_{Z^0}^2$. The limits of integral over $y$ and $p_T$ are
\begin{eqnarray}
-y^{max}(p_T)\leq y \leq y^{max}(p_T), \hspace{1.5cm}  0\leq p_T
\leq p_T^{max},
\end{eqnarray}
with
\begin{eqnarray}
&& y^{max}(p_T)={\rm arccosh}\bigg(\frac{S+ m_{Z^0}^2-
m_{A^0}^2}{2\sqrt{S (p_T^2 +m_{Z^0}^2)}}\bigg), \nonumber
\\ && p_T^{max}=\frac{1}{2\sqrt{S}} \sqrt{(S
+m_{Z^0}^2 -m_{A^0}^2)^2 -4m_{Z^0}^2S} \ .
\end{eqnarray}
The differential distribution with respect to $p_T$ and $y$ of
$A^0$ is similar to the one of $Z^0$. The differential
distribution with respect to the invariant mass $M_{AZ}$ is given
by
\begin{eqnarray}
\frac{d\sigma}{dM_{AZ}} =
\frac{2M_{AZ}}{S}\sum_{\alpha,\beta}\frac{d\cal{L}_{AZ}^{\alpha\beta}}{d\tau}\hat{\sigma}_{\alpha\beta}(\tau
S),
\end{eqnarray}
where $\Large{\frac{d\cal{L}_{AZ}^{\alpha\beta}}{d\tau}}$ is the
parton luminosity, defined as:
\begin{eqnarray}
\frac{d\cal{L}_{AZ}^{\alpha\beta}}{d\tau}=\int^1_\tau
\frac{dx}{x}\bigg[G_{\alpha/p}(x,\mu_f)G_{\beta/p}(\tau/x,\mu_f)
\bigg],
\end{eqnarray}
with
\begin{eqnarray}
&&M_{AZ}\equiv
\sqrt{(E_3+E_4)^2-(\overrightarrow{p_3}+\overrightarrow{p_4})^2} \/ \geq (m_{A^0}+m_{Z^0}),\\
&&\tau\equiv M^2_{AZ}/S.
\end{eqnarray}

\section{Numerical Results}

In the numerical calculations, we used the following set of
SM parameters\cite{SM}:
\begin{eqnarray}
&& \alpha_{ew}(m_W)=1/128, \ m_W=80.419 \, {\rm GeV},
\ m_Z=91.1882 \,{\rm GeV} \, ,\nonumber\\
&& m_t=178 \, {\rm GeV}, \ \alpha_s(M_Z)=0.118.
\end{eqnarray}
The running QCD coupling $\alpha_s(Q)$ is evaluated at the two-loop
order~\cite{runningalphas}, and the CTEQ6M PDFs \cite{CTEQ} is used
throughout this paper to calculate various cross sections, either at the
LO or NLO.
As to the Yukawa coupling of the bottom quark, we shall first use the
$\overline{\rm MS}$ bottom quark mass, $m_b(m_b)=4.25$\,GeV, to
evaluate the event rate, then compare it with the one calculated
using the QCD improved running mass to reduce the higher order QCD
radiative corrections, therefore improve the perturbative
calculations. The QCD improved running mass $m_b(Q)$, evaluated by
the NLO formula \cite{runningmb}, is:
\begin{equation}
m_b(Q)=U_6(Q,m_t)U_5(m_t,m_b)m_b(m_b) \, ,
\end{equation}
where the evolution factor $U_f$ is
\begin{eqnarray}
U_f(Q_2,Q_1)=\bigg(\frac{\alpha_s(Q_2)}{\alpha_s(Q_1)}\bigg)^{d^{(f)}}
\bigg[1+\frac{\alpha_s(Q_1)-\alpha_s(Q_2)}{4\pi}J^{(f)}\bigg], \nonumber \\
d^{(f)}=\frac{12}{33-2f}, \hspace{1.0cm}
J^{(f)}=-\frac{8982-504f+40f^2}{3(33-2f)^2} \, ,
\end{eqnarray}
and $f$ is the number of the active light quarks.
For comparison, we list the QCD improved
running bottom quark mass in Table~\ref{tc}
for various energy scale $Q$.

For large $\tan \beta$, the SUSY threshold correction to the bottom
quark Yukawa couplings could be large, and it can be resummed by making the
following replacement in the tree-level couplings
to improve the perturbation calculations~\cite{runningmb}:
\begin{eqnarray}
&& m_b(Q) \ \ \rightarrow \ \ \frac{m_b(Q)}{1+\Delta
m_b(Q=M_{SUSY})}, \label{deltamb}
\\
&& \Delta
m_b=\frac{2\alpha_s(Q=M_{SUSY})}{3\pi}M_{\tilde{g}}\mu\tan\beta
I(m_{\tilde{b}_1},m_{\tilde{b}_2},M_{\tilde{g}})
+\frac{h_t^2}{16\pi^2}\mu A_t\tan\beta
I(m_{\tilde{t}_1},m_{\tilde{t}_2},\mu) \nonumber \\
&& \hspace{1.0cm} -\frac{g^2}{16\pi^2}\mu M_2\tan\beta
\sum_{i=1}^2 \bigg[(R^{\tilde{t}}_{i1})^2
I(m_{\tilde{t}_i},M_2,\mu) + \frac{1}{2}(R^{\tilde{b}}_{i1})^2
I(m_{\tilde{b}_i},M_2,\mu)\bigg] \label{deltamb1} \, ,
\end{eqnarray}
where
\begin{eqnarray}
I(a,b,c)=\frac{1}{(a^2-b^2)(b^2-c^2)(a^2-c^2)}
(a^2b^2\log\frac{a^2}{b^2} +b^2c^2\log\frac{b^2}{c^2}
+c^2a^2\log\frac{c^2}{a^2}) \, ,
\end{eqnarray}
\begin{eqnarray}
h_t=\frac{gm_t}{\sqrt{2}m_W\sin{\beta}},
\end{eqnarray}
and $R^{\tilde{t}}$ and $R^{\tilde{b}}$ are the rotation matrices
for defining the mass eigenstates of ${\tilde{t}}$ and
${\tilde{b}}$, respectively. We set $M_{SUSY}$ in $\Delta m_b$ to
$m_{\tilde{g}}$ in our numerical calculations. Needless to say
that when using the running bottom quark Yukawa coupling to
evaluate cross sections, we shall subtract the corresponding
(SUSY-)QCD corrections at the order $\alpha_s$ from the
renormalization constant $\delta m_b$ to avoid double counting in
perturbative expansion of the strong coupling constant.

The values of the MSSM parameters taken in our numerical
calculations were constrained within the minimal supergravity
scenario (mSUGRA)~\cite{msugra}, in which there are only five free
input parameters at the grand unification (GUT) scale. They are
$m_{\frac{1}{2}},m_0,A_0,\tan\beta$ and the sign of $\mu$, where
$m_{\frac{1}{2}},m_0,A_0$ are, respectively, the universal gaugino
mass, scalar mass and the trilinear soft breaking parameter in the
superpotential. Given those parameters, all the MSSM parameters at
the weak scale are determined in the mSUGRA scenario by using the
program package SUSPECT 2.3 \cite{suspect}. In particular, we used
the running Higgs masses at the $m_{Z}$ scale, defined in the
modified dimensional reduction ($\overline{DR}$) scheme, which
have included the full one-loop corrections, as well as the
two-loop corrections controlled by the strong gauge coupling and
the Yukawa couplings of the third generation fermions
\cite{suspect,susyhiggs}. In our numerical calculations, we used
the two-loop renormalization group equations (RGEs) presented in
that program for calculating all the gauge couplings, the (third
generation) Yukawa couplings and the gaugino masses, while using
one-loop RGE for the other supersymmetric parameters. In the
following, we shall present our numerical studies based on the
five sets of SUSY input parameters listed in Table~\ref{ta}, which
are consistent with all the existing experiment data~\cite{SM}. We
will also vary $\tan\beta$, $m_0$ and $A_0$ to examine their
effects to various cross sections. For completeness, we also show
the relevant SUSY output parameters in Table~\ref{tb}. The QCD
plus SUSY-QCD and SUSY-EW improved bottom quark running mass are
listed in Table\ref{td}, which should be compared with those given
in Table~\ref{tc}, in which only QCD running effect is included.
For comparison, the QCD plus SUSY-QCD improved bottom quark running mass
are separately listed in Table\ref{te}.

\begin{table}[t!]
\begin{tabular}{|c|c|c|c|}
  \hline
   Q (GeV) & \ \ 250 \ \ \ & \ \ 500 \ \ \ & 750 \\
  \hline
  $m_b(Q)$ (GeV) \ \  & \ \ 2.68 \ \ & \ \ 2.55 \ \  & \ \ 2.49 \ \ \\
  \hline
 \end{tabular}\caption{The QCD improved running bottom quark
 mass, evaluated at $Q=250, 500$, and $750$\,GeV.
 The ${\overline{\rm MS}}$ bottom quark mass is taken to be
 $m_b(m_b)=4.25$\,GeV.
 }\label{tc}
  \end{table}
 \begin{table}[h!]
\begin{tabular}{|c|c|c|c|c|c|}
  \hline
set number  & $m_0$(GeV) \ \ & $m_{\frac{1}{2}}$(GeV) & $A_0$(GeV) \ \ & $\tan\beta$ & sign$(\mu)$  \\
  \hline
1 & 150  & 180 & 300 & 40 & + \\
  \hline
2 & 150  & 400 & 300 & 40 & + \\
  \hline
3 & 200  & 160  & 100 & 40 & - \\
  \hline
4 & 250  & 160  & 100 & 40 & - \\
 \hline
5 & 400 & 160  & 100 & 40 & - \\
  \hline
\end{tabular}\caption{ Five sets of SUSY input parameters studied in
this paper, within the mSUGRA scenario.
}\label{ta}
  \end{table}
\begin{table}[h!]
\begin{tabular}{|c|c|c|c|c|c|c|c|c|c|c|}
  \hline
 & $m_{\tilde{b}_{1(2)}}$(GeV) &
$m_{\tilde{t}_{1(2)}}$(GeV) &
  $m_{\tilde{g}}$(GeV) & $m_{A^0(h^0,H^0)}$(GeV) & $A_{t(b)}$(GeV) & $\mu$(GeV)& $\alpha$ & $\cos\theta_{\tilde{t}(\tilde{b})}$ \\
  \hline
1 & 374.6(429.1)  & 339.7(457.7) & 457.0 & 223.8(107.5,223.9) \ \ & -256.5(-275.8) & 235.3  & -0.032 & 0.97(0.74)\\
  \hline
2 & 764.3(822.0)  & 673.6(833.8) & 940.0 & 458.3(115.5,458.3) \ \ & -607.5(-750.9) & 498.1  & -0.027 & 0.47(0.71)\\
  \hline
3 & 314.3(395.1)  & 305.5(425.1) & 416.8 & 133.7(106.7,134.1) \ \ & -263.9(-303.9) & -224.6 & -0.143 & 0.62(-0.69)\ \ \\
  \hline
4 & 330.8(408.6)  & 317.0(434.3) & 419.9 & 155.0(107.2,155.3) \ \ & -262.7(-303.6) & -228.8 & -0.086 & 0.60(0.71)\\
 \hline
5 & 396.1(467.5)  & 363.9(476.1) & 431.9 & 233.0(108.4,233.2) \ \ & -261.0(-304.9) & -249.4 & -0.043 & 0.54(0.79)\\
  \hline
\end{tabular}\caption{The SUSY output parameters used in our
numerical calculations, corresponding to the five sets of SUSY input
parameters listed in
Table~\ref{ta}.}\label{tb}
  \end{table}
\begin{table}[h!]
\begin{tabular}{|c|c|c|c|c|c|c|c|}
  \hline
set number  & 1 & 2 &
  3 & 4 & 5 \\
  \hline
$m_b(Q=250, M_{SUSY}=m_{\tilde{g}})$(GeV) \ \ & \ \ 2.35 \ \ & \ \ 2.41 \ \ & \ \ 3.18 \ \ & \ \ 3.16 \ \ & \ \ 3.10 \ \ \\
 \hline
$m_b(Q=500, M_{SUSY}=m_{\tilde{g}})$(GeV) \ \ & \ \ 2.24 \ \ & \ \ 2.29 \ \ & \ \ 3.03 \ \ & \ \ 3.01 \ \ & \ \ 2.96 \ \ \\
 \hline
$m_b(Q=750, M_{SUSY}=m_{\tilde{g}})$(GeV) \ \ & \ \ 2.18 \ \ & \ \ 2.23 \ \ & \ \ 2.95 \ \ & \ \ 2.93 \ \ & \ \ 2.88 \ \ \\
  \hline\end{tabular}\caption{The QCD plus SUSY-QCD and SUSY-EW
  improved bottom quark running mass for the five sets of SUSY inputs listed in
Table~\ref{ta}, evaluated at $Q=250,500$ and
$750$\,GeV.}\label{td}
  \end{table}
\begin{table}[h!]
\begin{tabular}{|c|c|c|c|c|c|c|c|}
  \hline
set number  & 1 & 2 &
  3 & 4 & 5 \\
  \hline
$m_b(Q=250, M_{SUSY}=m_{\tilde{g}})$(GeV) \ \ & \ \ 2.15 \ \ & \ \ 2.14 \ \ & \ \ 3.71 \ \ & \ \ 3.66 \ \ & \ \ 3.53 \ \ \\
 \hline
$m_b(Q=500, M_{SUSY}=m_{\tilde{g}})$(GeV) \ \ & \ \ 2.04 \ \ & \ \ 2.04 \ \ & \ \ 3.53 \ \ & \ \ 3.49 \ \ & \ \ 3.36 \ \ \\
 \hline
$m_b(Q=750, M_{SUSY}=m_{\tilde{g}})$(GeV) \ \ & \ \ 1.99 \ \ & \ \ 1.98 \ \ & \ \ 3.44 \ \ & \ \ 3.39 \ \ & \ \ 3.27 \ \ \\
  \hline\end{tabular}\caption{The QCD plus SUSY-QCD
  improved bottom quark running mass for the five sets of SUSY inputs listed in
Table~\ref{ta}, evaluated at $M_{SUSY}=m_{\tilde{g}}$, and $Q=250,
500$, and $750$\,GeV.}\label{te}
  \end{table}
As for the renormalization and factorization scales, we always
chose $\mu_r=m_{\rm av}\equiv(m_{A^0}+m_{Z^0})/2$ and
$\mu_f=m_{\rm av}$, unless specified otherwise.

\subsection{LO total cross section}

 In Fig.~6 and Fig.~7, we first
compare the LO total cross sections of $pp\rightarrow A^0Z^0$ via
$b\bar{b}$ annihilation with the ones via gluon fusion and
Drell-Yan processes, respectively. Here, we use the $\overline{\rm
MS}$ bottom quark mass $m_b(m_b)=4.25$\,GeV, without including the
effect from QCD running.
 Our numerical results are different from
the ones presented in Ref.~\cite{AZLO}, because the
updated SUSY parameters are used instead of the earlier input parameters
used in Ref.~\cite{AZLO} which have already been ruled out
by recent experiments.
As shown in Figs.~6 and ~7, the LO total cross sections via $b\bar{b}$
annihilation and Drell-Yan processes increase with
$\tan\beta$, while the ones via gluon fusion process are
relatively larger for low and high values of $\tan\beta$,
 but become smaller for intermediate values of
$\tan\beta$. Moreover, all the LO rates decrease when
 $m_{A^0}$ increases. Figs.~6 and ~7 also show that in most of the
chosen parameter range, $b\bar{b}$ contributions are much larger
than the ones from gluon fusion and Drell-Yan processes, especially
for large $\tan\beta$ and small $m_{A^0}$, where the total cross
sections from the $b\bar{b}$ contributions can reach a few hundred
fb.
\subsection{Cutoff dependence}

In Fig.~8, we show the dependence of the NLO QCD predictions on
the two arbitrary theoretical cutoff scales $\delta_s$ and
$\delta_c$, introduced in the two-cutoff phase space slicing
method, where we have set $\delta_c=\delta_s/50$ to simplify the
study and used QCD plus SUSY improved bottom quark Yukawa coupling.
The NLO total cross section can be separated into two classes of
contributions. One is the $2 \to 2$ rate contributed by the Born
level, and the ${\cal{O}}(\alpha_s)$ virtual, soft and hard
collinear real emission corrections,
denoted as $\hat{\sigma}^{B}$, $\hat{\sigma}^{V}$, $\hat{\sigma}^{S}$
and $\sigma^{coll}$ in Eq.~(\ref{t}).
Another is the $2 \to 3$ rate
contributed by the ${\cal{O}}(\alpha_s)$ hard non-collinear real
emission corrections, denoted as
$\hat{\sigma}^{\overline{HC}}$ and $\hat{\sigma}^{\overline{C}}$
in Eq.~(\ref{t}).
 As noted in the previous section, the $2 \to
2$ and $2 \to 3$ rates depend individually on $\delta_s$ and
$\delta_c$, but their sum should not depend on any of the
theoretical cutoff scales. This is clearly illustrated in Fig.~8
for two different sets of SUSY parameters. We find that
$\sigma_{NLO}$ is almost unchanged for $\delta_s$ between $5\times
10^{-5}$ and $10^{-2}$, which is about 200\,fb and 28\,fb,
respectively for the two different sets of SUSY parameters.
Therefore, we take $\delta_s=10^{-4}$ and $\delta_c=\delta_s/50$
in the numerical calculations below.

\subsection{$m_{A^0}$ dependence}

In Fig.~9, we show the total cross sections of $pp\rightarrow
A^0Z^0$ at the LHC as a function of $m_{A^0}$ for
$\tan\beta=10$ and $40$, respectively, assuming
$m_{\frac{1}{2}}=160$\,GeV, $A_0=100$\,GeV, and $\mu<0$.
We considered the
LO total cross sections in three different cases, i.e. using (I)
$\overline{\rm MS}$ bottom quark mass at the scale $m_b$,
(II) QCD improved bottom quark running mass
at the scale $m_{A^0}$, and
(III) QCD plus SUSY improved bottom quark running mass
at the scale $m_{A^0}$,
respectively. We also considered the NLO total cross sections for
the cases of (II) and (III). Fig.~9 shows that the LO
and NLO total cross sections get smaller with the increasing
$m_{A^0}$, and the results for $\tan\beta=10$ in Fig.~9(2) are much
smaller than the ones for $\tan\beta=40$ in Fig.~9(1). For small
$m_{A^0}$ ($<160$\,GeV) the LO total cross sections in Fig.~9(1) can be
larger than $100$\,fb. The contributions
from the QCD running $m_b$ mass
effects and the SUSY improved  $m_b$ corrections are
significant, for example, in Fig.~9(1) when $m_{A^0}\simeq 155$\,GeV
and $\tan\beta=40$, the LO total cross sections are about 270\,fb,
120\,fb and 185\,fb for the three cases, respectively. Moreover, Fig.~9
shows that the NLO QCD corrections can either enhance or suppress
 the total rate, and the ${\cal O}(\alpha_s)$ contribution is
 in general a few tens percent of the total rate, as described below.
Define the K-factor as the ratio of the NLO to LO
total cross sections, calculated using the CTEQ6M PDFs.
We shown in Fig.~10 the dependence of the K factor
on $m_{A^0}$ for $A^0Z^0$ production, based on
the results of case (II) in Fig.~9. Namely, the QCD improved
bottom quark running mass is used for calculating the total cross
section at the LO and the NLO. Fig.~10 shows that in general the K
factor becomes smaller with the increasing $m_{A^0}$. For example,
the curve (a) in Fig.~10(1) shows that when $m_{A^0}$ varies from
108\,GeV to 900\,GeV, the K factor varies from 1.72 to 0.82, and
the curve (a) in Fig.~10(2) shows that when $m_{A^0}$ varies from
235\,GeV to 860\,GeV, the K factor varies from 0.91 to 0.68. The
contributions to the K factors, shown as curve (a), in both
Fig.~10(1) and Fig.~10(2) come from the pure QCD corrections,
shown as curve (b), and SUSY QCD corrections, shown as curve (c).
The former includes both the virtual and real emission
contributions originated from pure QCD corrections, while the
latter consists of only virtual corrections. As expected, the
$K$-factor contributed by the pure QCD corrections is under
controlled, of a few tens percent, when the QCD improved bottom
quark running mass is used to evaluate the Yukawa coupling of
the bottom quark. On the other hand, the SUSY QCD
corrections could become large as $m_{A^0}$ decreases, especially
for large $\tan\beta$. For example, in Fig.~10(1), for
$\tan\beta=40$, when $m_{A^0}\simeq108$\,GeV, the $K$ factor of
SUSY QCD corrections is about 0.8 which dominates the overall K
factor. Hence, to improve the convergence of the perturbation
calculations in the case of large $\tan\beta$, we could use the
SUSY improved bottom quark running mass to evaluate the Yukawa
coupling of bottom quark.
More on SUSY QCD corrections will be
discussed below.
We have also examined the contributions
from the box diagrams shown in Fig.~3. The pure QCD box diagram
contribution, arising from Fig.~3(a) and Fig.~3(c), is ultraviolet
finite but not infrared finite. For $\tan\beta=40$ the finite part
of the pure QCD box diagram contribution becomes more important
for large $m_{A^0}$, and its effect is to decreases the total
rate. On the contrary, the SUSY QCD box diagram contribution,
arising from Fig.~3(b) and Fig.~3(d), is free of any singularity,
and is small numerically.

Fig.~11 shows the dependence of the K factors on $m_{A^0}$ for
$A^0Z^0$ production, based on the results of case (III) in
Fig.~9. Namely, the QCD plus SUSY improved bottom quark running
Yukawa coupling is used for calculating the total cross
section at the LO and
the NLO. Generally, the K factor decreases with $m_{A^0}$. For
example, for $\tan\beta=40$,
 when $m_{A^0}$ varies from 108\,GeV to 900\,GeV, the K
factor corresponding to curve (a) ranges from 0.98 to 0.61, which
contains two parts: the pure QCD corrections, shown as curve (b),
and SUSY QCD corrections, shown as curve (c). As compared to the
results in Fig.~10(1), we find that the SUSY QCD correction, shown
as curve (c), has been largely suppressed. For instance, the K
factor of SUSY QCD corrections drops from 0.8, in Fig.~10(1), to
0.05, in Fig.~11, for $m_{A^0}\simeq108$\,GeV, while the other
SUSY parameters are identical in both calculations. This is
because using the SUSY improved running $m_b$ to evaluate the LO
cross section, we have already included the dominant NLO SUSY QCD
corrections.
 Therefore, we shall use the QCD plus SUSY improved bottom
quark running mass in the following numerical analysis
for both the LO and NLO calculations, unless
specified otherwise.

\subsection{SUSY QCD corrections in heavy mass limit}

It is instructive to examine the results of Figs.~10(1) and
Fig.~11
 in the heavy mass
limit, where all the SUSY mass parameters except $m_{A^0}$ are of
the same size and tend to be heavy, i.e. $M_{\tilde{Q}}$ $=$
$M_{\tilde{D}}$ $=$ $\mu$ $=$ $A_b$ $=$ $M_{\tilde{g}}$ $\equiv
M_{SUSY}$ $ \gg m_Z$.
In the heavy mass limit, the SUSY QCD box diagram contribution,
arising from
Fig.~3(b) and Fig.~3(d), is suppressed by powers of $M_{SUSY}$ and can
be neglected. This is confirmed by our numerical calculation which shows
that the SUSY QCD box contribution is generally below $0.1\%$ of the
total rate.
Hence, we shall examine the effect of
SUSY QCD corrections in the heavy mass limit
to the virtual diagrams shown in Figs.~2(a)--(g),
and compare the analytical result with our numerical
calculations.

Since our aim is to examine the NLO SUSY QCD effect in this part
of study, we shall use the LO bottom quark Yukawa coupling (with
$m_b=4.25$\,GeV) to evaluate the relevant tree level vertices.
Keeping only terms at ${\cal O}(\alpha_s)$ that are not suppressed
by negative powers of heavy mass $M_{SUSY}$ in the heavy mass
limit, the one loop SUSY QCD correction to the individual diagram
in Fig.~2 yields the following corrections. After stripping off
the Born level matrix element (including all the vertex and
propagator factors), the multiplicative factor of the s-channel
diagram with the $h^0$ propagator, cf. Fig.~2(a), is given by
\begin{eqnarray}
F_{(a)h^0}=-\frac{g_s^2}{12\pi^2}(1+\cot\alpha) \, \label{haber}
\end{eqnarray}
where $\alpha$ is the mixing angle of the two CP-even Higgs
bosons~\cite{mssm}.
Note that Eq.~(\ref{haber}) is in agreement with the one shown in
Ref.~\cite{anal}.
Similarly, the multiplicative factor of the s-channel diagram with the
$H^0$ propagator, cf. Fig.~2(a), is given by
\begin{eqnarray}
F_{(a)H^0}=-\frac{g_s^2}{12\pi^2}(1-\tan\alpha) \,.
\end{eqnarray}
The multiplicative factor of the t and u-channel diagrams, cf.
Fig.~2(c) or (d), is given by
\begin{eqnarray}
F_{(c)A^0}=F_{(d)A^0}=-\frac{g_s^2}{12\pi^2}(1+\cot\beta)\,.
\end{eqnarray}
The multiplicative factor for the sum of Figs.~2(b) and 2(g) is
zero. This is because after adding the wavefunction
renormalization factor for the external bottom quark line, the
renormalized $Zb{\bar b}$ vertex vanishes in the heavy mass limit.
(Again, we have dropped any term that is suppressed by negative
powers of the heavy mass scale $M_{SUSY}$.) Similarly, the
multiplicative factor for the sum of Figs.~2(e) and 2(f) is zero.

Given the above multiplicative factors, we can calculate the SUSY
QCD correction to the total cross section for $pp\rightarrow
A^0Z^0$ production at the LHC, and compare it with the complete
numerical calculation described in Sec. III. The results in the heavy
mass limit are shown
in Fig.~12, which show that the agreement becomes better for
larger value of $M_{SUSY}$.
Hence, this provides a consistent check on our complete numerical
calculations.

\subsection{$\tan\beta$ dependence}

In Figs. 13~(1) and 13~(2), the total cross sections for
$pp\rightarrow A^0Z^0$ at the LHC are plotted as a function of
$\tan\beta$ for two representative values of $m_{\frac{1}{2}}$ and
$m_0$, respectively. In Fig.~13(2), when $\tan\beta$ ranges between
4 and 40, $m_{A^0}$ varies from 330\,GeV to 223\,GeV,
 and from 660\,GeV
to 458\,GeV for $m_{\frac{1}{2}}=180$\,GeV and $400$\,GeV, respectively.
From Fig.~13(2) we can clearly see that the LO and NLO total cross
sections are enhanced with the increasing $\tan\beta$ and decreased
with the increasing $m_{\frac{1}{2}}$. For large $\tan\beta$
($>40$) and $m_{\frac{1}{2}}=180$\,GeV, the LO and NLO total cross
sections can be over 30\,fb. The features in Fig.~13(1) are similar
to the ones in Fig.~13(2), but in general the total cross sections
are larger than later. For example, for large $\tan\beta(>40)$ and
$m_{0}=200$\,GeV, both of the LO and NLO total cross sections can
reach about hundreds of fb.

Fig.~14 shows the dependence of the K factors on $\tan\beta$,
based on the results in Fig.~13, where the K factor increases with
the increasing $\tan\beta$. For the results of Fig.~13(1), the K
factor varies from 0.69 to 0.90 and from 0.65 to 0.92 for
$m_0=200$\,GeV and $400$\,GeV, respectively. For the results of
Fig.~13(2), the K factor varies from 0.70 to 0.74 and from 0.62 to
0.63 for $m_{\frac{1}{2}}=180$\,GeV and $400$\,GeV, respectively.

\subsection{$\mu_r/\mu_f$ dependence}

Fig.~\ref{scale} shows the dependence of the total cross sections
for $pp\rightarrow A^0Z^0$ production at the LHC on the
renormalization scale ($\mu_r$) and the factorization scale
($\mu_f$), with $\mu_r=\mu_f$. The case (1) is for $\mu<0$, and the
case (2) is for $\mu>0$.
In both cases, the scale dependence of the NLO
total cross section is smaller than that of the LO cross section.
For example, the LO cross sections vary from 65\,fb to 261\,fb and
25\,fb to 45\,fb when $\mu_r=\mu_f$ ranges between $0.1\,m_{\rm
av}$ and $10\,m_{\rm av}$, while the NLO ones vary from 230\,fb to
232\,fb and 38\,fb to 39\,fb, in the case (1) and (2),
respectively. Here, the QCD plus
SUSY improved bottom quark Yukawa coupling is used.
For comparison, we also show the results of other two calculations.
The case (3) is similar to the case (1), but in (3) the pure QCD running
bottom quark mass is used instead. The case (4) is
similar to the case (1), but in (4) the contribution from the SUSY-EW
correction in the running bottom quark Yukawa coupling
 is not included, namely,
only the pure QCD and SUYSY-QCD corrections are included.

To further investigate the scale dependence in case (1), with $\mu
<0$, we study the scale dependence of the total cross section on
the renormalization scale $(\mu_r)$ and the factorization scale
$(\mu_f)$ seperatedly in Fig.~\ref{scale2}. Here, the QCD plus
SUSY improved bottom quark Yukawa coupling is used. We find that
in either case, whether we fixed $\mu_r$ and let $\mu_f$ vary, or
vice versa, the NLO rate is less dependent on the scale than the
LO rate.

Hence, when applying the usual prescription to estimate the scale
dependence, i.e. varying the scale around $m_{av}$ by a factor of
2, the NLO cross sections vary by around 10\% to 20\%, cf.
Figs.~\ref{scale} and \ref{scale2},

\subsection{PDF uncertainty}

To estimate the uncertainties in the total cross sections due to
the uncertainty of PDFs, we take the 41 sets of CTEQ6.1 PDFs to
calculate the LO and NLO rates~\cite{61cteq}. As shown in
Fig.~\ref{cteq}, the LO result of using the CTEQ6M PDF lies
between the maximum ($\sigma_{max}$) and
minimum ($\sigma_{min}$)  LO rates. The NLO total cross
sections are then calculated using three different PDF sets, one
of which is CTEQ6M, the other two are the ones that give the
maximum and minimum LO rates, respectively. The total cross
sections for $pp\rightarrow A^0Z^0$ production at the LHC, as a
function of the trilinear coupling $A_0$, for the above mentioned
PDFs are shown in Fig.~\ref{cteq}, where we have used the QCD
running mass to evaluate the bottom quark Yukawa coupling. It
turns out that the PDF uncertainties (defined here as $\pm
(\sigma_{max}-\sigma_{min}))/(\sigma_{max}+\sigma_{min})$) in the
LO and NLO total cross sections are about the same, when the QCD
running $m_b$ is used. For example, when $A_0=100$\,GeV, the PDF
uncertainties are $\pm 2.9\%$ at the LO, and $\pm 3.0\%$ at the
NLO, respectively.

Fig.~\ref{PDFU} shows the PDF uncertainties (defined here as the
Eq.~(3) in Ref.~\cite{PDFUU}) in the LO and NLO total cross
sections for $pp\rightarrow A^0Z^0$ production
at the LHC, as a function of
$m_{A^0}$. Here, we also used the QCD running mass to
evaluate the bottom quark Yukawa coupling. It turns out
that the NLO rate has a slightly larger uncertainty than the LO rate
due to the PDF uncertainties, especially at large $m_{A^0}$.
Also, the uncertainty in the total cross section
at the LHC increases as $m_{A^0}$ increases.

\subsection{Differential cross sections}

Fig.~\ref{pt1} shows the differential cross section as a
function of the transverse momentum $p_T$ of $Z^0$ and $A^0$ in
the associated production of the $A^0Z^0$ pairs at the LHC.
We find that the NLO QCD correction could change the shape of
transverse momentum distribution. The NLO QCD correction
enhances the LO differential cross section in low and
high $p_T$ region, but reduces in medium $p_T$ region.

Fig.~\ref{pt2} shows the differential cross section as a
function of the invariant mass $M_{A^0Z^0}$ of the $A^0Z^0$
pairs produced at the LHC. The NLO QCD
corrections reduce the LO differential cross sections
more in the medium values of $M_{A^0Z^0}$, and much less in low or high
values of $M_{A^0Z^0}$.

\section{Conclusions}

In conclusion, we have calculated the complete NLO QCD corrections
to the inclusive total cross sections of the $A^0Z^0$ pairs
produced at the LHC in the MSSM. We have preformed the
calculations using both the DREG and DRED schemes, and found that
the NLO total cross sections in the above two schemes are the
same, which provides a cross check to our calculations. Our
results show that the LO total cross sections are a few tens fb in
most of the SUSY parameter space, and can exceed 100\,fb for
$m_{A^0}$ below 160\,GeV with large $\tan\beta$($\gtrsim 40$). The
NLO correction can either enhance or reduce the total cross
sections, but it generally efficiently reduces the dependence of
the total cross sections on the renormalization/factorization
scale. For small $m_{A^0}$ and large $\tan\beta$, the $K$-factor
of SUSY QCD corrections could become large, and using the QCD plus
SUSY improved Yukawa coupling in the calculation could reduce the
size of the overall $K$-factor. We have also examined the
uncertainty in total cross sections due to the PDF uncertainties,
and found that the uncertainty in NLO cross sections is slightly
larger than that in LO ones, especially at large $m_{A^0}$.
Finally, we also examined a few differential distributions and
found that the NLO QCD corrections could change the shape of
transverse momentum and invariant mass distributions.

\begin{acknowledgments}
We thank Qing-Hong Cao and Tao Han for useful discussion. This
work was supported in part by the National Natural Science
Foundation of China and Specialized Research Fund for the Doctoral
Program of Higher Education.
The work of CPY was supported in part by the USA NSF grant
PHY-0244919.
\end{acknowledgments}

\section*{Appendix A}
In this appendix, we give the relevant Feynman rules. \\
1. $h^0(H^0)-b-\bar{b}: \ \ A_{1(2)}m_b$
\begin{eqnarray}
&& A_1=\frac{igs_\alpha}{2m_w\cos\beta},\ \
A_2=\frac{-igc_\alpha}{2m_w\cos\beta}, \nonumber\end{eqnarray}
where $\alpha$ is the mixing angle in the CP even neutral Higgs boson
sector~\cite{mssm}. Here, we use the abbreviations $s_\alpha=\sin\alpha$
and
$c_\alpha=\cos\alpha$.\\
 2. $A^0-b-\bar{b}: \ \ A_{3}m_b\gamma_5$
\begin{eqnarray}
&& A_3=\frac{-g\tan\beta}{2m_w}. \nonumber\end{eqnarray}\\
3. $h^0(H^0)-Z^0-A^0: \ \ F_{1(2)}(p_{h^0(H^0)}+p_{A^0})^{\mu}$
\begin{eqnarray}
&& F_1=\frac{g\cos(\beta-\alpha)}{2\cos\theta_w},
 F_2=\frac{-g\sin(\beta-\alpha)}{2\cos\theta_w}, \nonumber\end{eqnarray}
 Here, we define the outgoing four-momenta of $h^0(H^0)$ and $A^0$
 to be negative and positive, respectively.
 \\
 4. $Z^0-b-\bar{b}: \ \ \gamma_{\mu}(C_V+C_A\gamma_5)$
\begin{eqnarray}
&&
C_V=\frac{-ig}{2\cos\theta_w}\bigg(-\frac{1}{2}+\frac{2}{3}sin^2\theta_w\bigg),
\ \ C_A=\frac{-ig}{4\cos\theta_w}. \nonumber\end{eqnarray}
 5.
$h^0(H^0,A^0)-\tilde{b}_\alpha-\tilde{b}_\beta: \ \
i[R^{\tilde{b}}\hat{G}^{\tilde{b}}_{1(2,3)}(R^{\tilde{b}})^T]_{\alpha\beta}$
\begin{eqnarray}
\hat{G}^{\tilde{b}}_1= \left(\begin{array}{cc}
\frac{gm_{Z}}{\cos\theta_w}(-\frac{1}{2}+\frac{1}{3}\sin\theta^2_w)\sin(\alpha+\beta)+\sqrt{2}m_bh_b
s_\alpha & \frac{1}{\sqrt{2}}h_b\left[A_b s_\alpha +\mu c_\alpha
\right]\\
\frac{1}{\sqrt{2}}h_b \left[A_b s_\alpha +\mu c_\alpha \right] &
\frac{gm_{Z}}{\cos\theta_w}(-\frac{1}{3}\sin\theta^2_w)\sin(\alpha+\beta)+\sqrt{2}m_bh_bs_\alpha
 \end{array} \right),\nonumber
\end{eqnarray}
\begin{eqnarray}
\hat{G}^{\tilde{b}}_2= \left(\begin{array}{cc}
\frac{gm_{Z}}{\cos\theta_w}(\frac{1}{2}-\frac{1}{3}\sin\theta^2_w)\cos(\alpha+\beta)-\sqrt{2}m_bh_b
c_\alpha & -\frac{1}{\sqrt{2}}h_b\left[A_b c_\alpha -\mu s_\alpha  \right] \\
-\frac{1}{\sqrt{2}}h_b\left[A_b c_\alpha -\mu s_\alpha \right] &
\frac{gm_{Z}}{\cos\theta_w}(\frac{1}{3}\sin\theta^2_w)\cos(\alpha+\beta)-\sqrt{2}m_bh_b
c_\alpha
\end{array} \right),\nonumber
\end{eqnarray}
\begin{eqnarray}
\hat{G}^{\tilde{b}}_3=i\frac{gm_b}{2m_W} \left(\begin{array}{cc} 0
& -A_b \tan\beta  -\mu \\
A_b \tan\beta +\mu & 0 \end{array} \right),\nonumber
\end{eqnarray}
with
$h_{b}=\frac{gm_{b}}{\sqrt2m_{W}\cos\beta}$.\\
6. $Z^0-\tilde{b}_\alpha-\tilde{b}_\beta: \ \
\frac{-ig}{cos\theta_w}T_Z(\alpha,\beta)
(p_{\tilde{b}_\alpha}+p_{\tilde{b}_\beta})^\mu$
\begin{eqnarray}
T_Z= \left(\begin{array}{cc}
-\frac{1}{2}\cos^2_{\theta_{\tilde{b}}}+\frac{1}{3}\sin^2_{\theta_w}
& \frac{1}{4}\sin2\theta_{\tilde{b}} \\
\frac{1}{4}\sin2\theta_{\tilde{b}} &
-\frac{1}{2}\sin^2_{\theta_{\tilde{b}}}+\frac{1}{3}\sin^2_{\theta_w}
\end{array} \right),\nonumber
\end{eqnarray}
where $p_{\tilde{b}_\alpha}$ and $p_{\tilde{b}_\beta}$ are the
four-momenta of $\tilde{b}_\alpha$ and $\tilde{b}_\beta$ in
direction of the charge flow.
\section*{Appendix B}

In this appendix, we collect the explicit expressions of the
nonzero form factors in Eq.~(\ref{fvertex}) and
Eq.~(\ref{fbox}).
Since
$\overline{\sum}M^0M^\dagger_{j=7,8,...12}=0$, only the form
factors of the first six matrix elements are presented here.
For
simplicity, we introduce the following abbreviations for the
Passarino-Veltman three-point integrals $C_{i(j)}$ and four-point
integrals $D_{i(j)}$, which are defined similar to
Ref.~\cite{denner} except that we take internal masses squared as
arguments:

\noindent $C^a_{i(j)}=C_{i(j)}(0,0,s,0,0,0)$,
\\
$C^b_{i(j)}=C_{i(j)}(m^2_{A^0},t,0,0,0,0)$,
\\
$C^c_{i(j)}=C_{i(j)}(m^2_Z,u,0,0,0,0)$,
\\
$C^d_{i(j)}=C_{i(j)}(m^2_{A^0},0,u,0,0,0)$,
\\
$C^e_{i(j)}=C_{i(j)}(m^2_{z},0,t,0,0,0)$,
\\
$C^f_{i(j)}=C_{i(j)}(t,0,m^2_{z},0,0,0)$,
\\
$C^g_{i(j)}=C_{i(j)}(u,0,m^2_{z},0,0,0)$,
\\
$C^h_{i(j)}=C_{i(j)}(u,0,m^2_{A^0},0,0,0)$,
\\
$C^i_{i(j)}=C_{i(j)}(m^2_Z,t,0,0,0,0)$,
\\
$C^j_{i(j)}=C_{i(j)}(u,m^2_{A^0},0,0,0,0)$,
\\
$C^k_{i(j)}=C_{i(j)}(m^2_{A^0},t,0,0,0,0)$,
\\
$C^l_{i(j)}=C_{i(j)}(0,m^2_{Z},u,0,0,0)$,
\\
$C^u_{i(j)}=C_{i(j)}(s,0,0,0,0,0)$,
\\
$C^v_{i(j)}=C_{i(j)}(0,u,m^2_{Z},0,0,0)$,
\\
$C^x_{i(j)}=C_{i(j)}(s,m^2_{A^0},m^2_{Z},0,0,0)$,
\\
$C^y_{i(j)}=C_{i(j)}(0,m^2_{A^0},t,0,0,0)$,
\\
$C^z_{i(j)}=C_{i(j)}(0,t,m^2_{Z},0,0,0)$,
\\
$C^m_{i(j)}(a,b)=C_{i(j)}(0,0,s,m^2_{\tilde{b}_a},m^2_{\tilde{g}},m^2_{\tilde{b}_b})$,
\\
$C^n_{i(j)}(a,b)=C_{i(j)}(m^2_{A^0},t,0,m^2_{\tilde{b}_b},m^2_{\tilde{b}_a},m^2_{\tilde{g}})$,
\\
$C^o_{i(j)}(a,b)=C_{i(j)}(0,u,m^2_{Z},m^2_{\tilde{b}_b},m^2_{\tilde{b}_a},m^2_{\tilde{g}})$,
\\
$C^p_{i(j)}(a,b)=C_{i(j)}(m^2_{A^0},0,u,m^2_{\tilde{b}_b},m^2_{\tilde{b}_a},m^2_{\tilde{g}})$,
\\
$C^q_{i(j)}(a,b)=C_{i(j)}(t,m^2_Z,0,m^2_{\tilde{g}},m^2_{\tilde{b}_b},m^2_{\tilde{b}_a})$,
\\
$C^r_{i(j)}(a,b)=C_{i(j)}(u,0,m^2_{A^0},m^2_{\tilde{b}_b},m^2_{\tilde{b}_a},m^2_{\tilde{g}})$,
\\
$C^s_{i(j)}(a,b)=C_{i(j)}(u,m^2_{A^0},0,m^2_{\tilde{g}},m^2_{\tilde{b}_b},m^2_{\tilde{b}_a})$,
\\
$C^t_{i(j)}(a,b)=C_{i(j)}(0,m^2_{A^0},t,m^2_{\tilde{g}},m^2_{\tilde{b}_b},m^2_{\tilde{b}_a})$,
\\
$D^a_{i(j)}=D_{i(j)}(s,0,t,m^2_{A^0},0,m^2_Z,0,0,0,0)$,
\\
$D^b_{i(j)}=D_{i(j)}(0,t,m^2_Z,s,m^2_{A^0},0,0,0,0,0)$,
\\
$D^c_{i(j)}=D_{i(j)}(s,0,u,m^2_Z,0,m^2_{A^0},0,0,0,0)$,
\\
$D^d_{i(j)}=D_{i(j)}(0,u,m^2_{A^0},s,m^2_z,0,0,0,0,0)$,
\\
$D^e_{i(j)}(a,b,l)=D_{i(j)}(s,0,t,m^2_{A^0},0,m_Z^2,m^2_{\tilde{b}_b},m^2_{\tilde{b}_a},m^2_{\tilde{g}},m^2_{\tilde{b}_l})$,
\\
$D^f_{i(j)}(a,b,l)=D_{i(j)}(0,0,m^2_{Z},m^2_{A^0},s,t,m^2_{\tilde{b}_b},m^2_{\tilde{g}},m^2_{\tilde{b}_a},m^2_{\tilde{b}_l})$,
\\
$D^h_{i(j)}(a,b,l)=D_{i(j)}(s,0,u,m^2_{Z},0,m_{A^0}^2,m^2_{\tilde{b}_b},m^2_{\tilde{b}_a},m^2_{\tilde{g}},m^2_{\tilde{b}_l})$.
\\
Many of the
above functions contain the soft and/or collinear singularities.
Since all the Passarino-Veltman integrals can be written as a
combination of the
scalar functions $A_0$, $B_0$, $C_0$ and $D_0$, we present here the
explicit expressions for the
$C_0$ and $D_0$ functions used in our calculations:
\begin{eqnarray}
&&C^a_0=C^u_0=\frac{C_\epsilon}{s}\bigg[\frac{1}{\epsilon^2}-\frac{\pi^2}{3}\bigg],\nonumber\\
&&
C^d_0=C^h_0=C^j_0=\frac{C_\epsilon}{u-m^2_{A^0}}\bigg[\frac{1}{\epsilon}\ln\bigg(\frac{-u}{m^2_{A^0}}\bigg)
 +\frac{1}{2}\ln^2\bigg(\frac{s}{m^2_{A^0}}\bigg)-\frac{1}{2}\ln^2\bigg(\frac{s}{-u}\bigg)-\frac{\pi^2}{2}\bigg],\nonumber\\
&&
C^c_0=C^g_0=C^l_0=C^v_0=\frac{C_\epsilon}{u-m_Z^2}\bigg[\frac{1}{\epsilon}\ln\bigg(\frac{-u}{m^2_Z}\bigg)
 +\frac{1}{2}\ln^2\bigg(\frac{s}{m_Z^2}\bigg)-\frac{1}{2}\ln^2\bigg(\frac{s}{-u}\bigg)-\frac{\pi^2}{2}\bigg],\nonumber\\
&&
C^e_0=C^f_0=C^i_0=C^y_0=\frac{C_\epsilon}{t-m_Z^2}\bigg[\frac{1}{\epsilon}\ln\bigg(\frac{-t}{m^2_Z}\bigg)
 +\frac{1}{2}\ln^2\bigg(\frac{s}{m_Z^2}\bigg)-\frac{1}{2}\ln^2(\frac{s}{-t})-\frac{\pi^2}{2}\bigg],
\nonumber\\
&&
C^b_0=C^k_0=C^x_0=\frac{C_\epsilon}{t-m^2_{A^0}}\bigg[\frac{1}{\epsilon}\ln\bigg(\frac{-t}{m^2_{A^0}}\bigg)
 +\frac{1}{2}\ln^2\bigg(\frac{s}{m^2_{A^0}}\bigg)-\frac{1}{2}\ln^2\bigg(\frac{s}{-t}\bigg)-\frac{\pi^2}{2}\bigg],\nonumber\\
&& D^a_0=D^b_0=
  \frac{C_\epsilon}{st}\bigg[\frac{1}{\epsilon^2}+\frac{2}{\epsilon}\ln\bigg(\frac{m_Zm_{A^0}}{-t}\bigg)+\frac{\pi^2}{3}\bigg]
  -2\frac{C_\epsilon}{st}\Bigg \{{\rm Li}\bigg(\frac{m^2_{A^0}-u}{s}\bigg)
  -{\rm Li}\bigg(\frac{s-m_Z^2}{s}\bigg) \nonumber\\&&
 \ \ \ \ \ \ \ -{\rm Li}\bigg[\frac{-st}{(s-m_Z^2)(m_Z^2-t)}\bigg]
  +{\rm Li}\bigg(\frac{-t}{s-m_Z^2}\bigg)+{\rm Li}\bigg(\frac{m^2_{A^0}}{m^2_{A^0}-t}\bigg)
  -\frac{1}{2}\ln^2\bigg[\frac{-st}{(s-m_Z^2)(m_Z^2-t)}\bigg]\nonumber\\&&
  \ \  \ \ \ \ \
  +\frac{1}{2}\ln^2\bigg(\frac{-t}{s-m_Z^2}\bigg)
  +\ln\bigg(\frac{m_Z^2-t}{s}\bigg)\ln\bigg(\frac{m^2_{A^0}-u}{t}\bigg)
  -\frac{1}{2}\ln\bigg(\frac{m_Z^2-t}{s}\bigg)\ln\bigg(\frac{sm^2_{A^0}}{t^2}\bigg)\nonumber\\&& \ \ \ \ \ \ \
  +\frac{1}{4}\ln^2\bigg(\frac{sm^2_{A^0}}{t^2}\bigg)
  +\frac{1}{4}\ln^2\bigg(\frac{m^2_{Z}}{s}\bigg)
  +\frac{1}{2}\ln\bigg(\frac{m^2_{A^0}}{s}\bigg)\ln\bigg(\frac{m_Z^2-t}{m^2_{Z}}\bigg)
  +\frac{1}{2}\ln^2\bigg(\frac{m^2_{A^0}}{m^2_{A^0}-t}\bigg)
  \Bigg \},\nonumber
 \\
&& D^c_0=D^d_0=D^a_0(t\leftrightarrow u,m^2_{A^0}\leftrightarrow
m_Z^2),\nonumber
  \end{eqnarray}
where
$C_\epsilon=(4\pi\mu^2_r/s)^\epsilon\Gamma(1-\epsilon)/\Gamma(1-2\epsilon)$.
\\
For diagrams(a)-(g) in Fig.~2, we get the form factors as
following, respectively,
\begin{eqnarray}
&&
f^a_1=\frac{-4\alpha_s}{3\pi(s-m^2_{h_0})(s-m^2_{H_0})}\{ism_b[A_1F_1(s-m^2_{H_0})
+A_2F_2(s-m^2_{h_0})]C_0^a\nonumber\\
&& \hspace{0.6cm}+F_1m_{\tilde g}(s-m^2_{H_0})\hat{G}^{\tilde
b}_1(a,b)R^{\tilde b}_{a,1}R^{\tilde b}_{b,2}C^m_0 +F_2m_{\tilde
g}(s-m^2_{h_0})\hat{G}^{\tilde b}_2(a,b)R^{\tilde
b}_{a,1}R^{\tilde b}_{b,2}C^m_0\}\nonumber
\\&&\hspace{0.6cm}+\frac{4\alpha_s\epsilon}{3\pi(s-m^2_{h_0})(s-m^2_{H_0})}[A_1F_1(s-m^2_{H_0})
+A_2F_2(s-m^2_{h_0})]B_0(s,0,0),\nonumber
\\
&& f^a_2=f^a_1,\nonumber
\\
&& f^a_3=\frac{4m_{\tilde g}\alpha_s(R^{\tilde b}_{a,2}R^{\tilde
b}_{b,1}-R^{\tilde b}_{a,1}R^{\tilde
b}_{b,2})}{3\pi(s-m^2_{h_0})(s-m^2_{H_0})}[F_1\hat{G}^{\tilde
b}_1(a,b)(s-m^2_{H_0})+F_2\hat{G}^{\tilde
b}_2(a,b)(s-m^2_{h_0})]C^m_0,\nonumber
\\
&& f^a_4=f^a_3,\nonumber
\\
&& f^b_1=\frac{4im_bA_3(C_V-C_A)\alpha_s}{3\pi
t}[(1-\epsilon)(2C^f_{00}-uC^f_{12}+(t-m_{A_0}^2)C^f_{11}+m^2_ZC^e_1,
\nonumber\\
&&\hspace{0.9cm}-(s-m^2_Z-m^2_{A_0})C^f_{12}+m^2_{A_0}C^f_{11}-B_0(t,0,0))+m_Z^2C^i_0+tC^f_1]
\nonumber\\
&&\hspace{0.9cm}-\frac{4m_bg\alpha_sA_3T_{Z}(a,b)R^{\tilde
b}_{a1}R^{\tilde b}_{b1}}{3\pi tcos\theta_{w
}}[2c^q_{00}+uC^q_2+uC^q_{22}+(t-m^2_{A_0})(C^q_1+C^q_{11}+C^q_2+C^q_{22})\nonumber\\
&&\hspace{0.9cm}
+(s-m^2_Z-m^2_{A_0})(C^q_2+C^q_{22})+m^2_{A_0}(C^q_1+C^q_{11}+C^q_2+C^q_{22})+tC^q_{12}],
\nonumber
\\&&f^b_3=\frac{-4im_bC_VA_3\alpha_s}{3\pi
t}\{(1-\epsilon)[2C^f_{00}-uC^f_{12}+(t-m_{A_0}^2)C^f_{11}+m^2_ZC^e_1
\nonumber\\
&&\hspace{0.9cm}-(s-m^2_Z-m^2_{A_0})C^f_{12}+m^2_{A_0}C^f_{11}-B_0(t,0,0)]+m_Z^2C^i_0+tC^f_1\}
\nonumber\\
&&\hspace{0.9cm}+\frac{4m_bg\alpha_sA_3T_{Z}(a,b)R^{\tilde
b}_{a1}R^{\tilde b}_{b1}}{3\pi tcos\theta_{w
}}[2C^q_{00}+uC^q_2+uC^q_{22}+(t-m^2_{A_0})(C^q_1+C^q_{11}+C^q_2+C^q_{22})\nonumber\\
&&\hspace{0.9cm}
+(s-m^2_Z-m^2_{A_0})(C^q_2+C^q_{22})+m^2_{A_0}(C^q_1+C^q_{11}+C^q_2+C^q_{22})+tC^q_{12}]
,\nonumber
\\&&f^b_5=\frac{-2im_b\alpha_sA_3(C_V-C_A)}{3\pi
t}\{(1-\epsilon)[2C^f_{00}+(t-m^2_{A_0})C^f_1+m_Z^2C^e_1\nonumber\\
&&\hspace{0.9cm}+(s-m^2_Z-m^2_{A_0})(C^e_1-C^f_2) +
m^2_{A_0}(2C^f_1-C^e_2)-B_0(t,0,0)]+m_Z^2C^i_0\}\nonumber\\
&&\hspace{0.9cm}+\frac{2gm_b\alpha_sA_3T_{Z}(ab)R^{\tilde
b}_{a1}R^{\tilde b}_{b1}}{3\pi cos\theta_{w }t}C^q_{00},\nonumber
\\&&f^b_6=\frac{4im_b\alpha_sC_VA_3}{3\pi
t}\{(1-\epsilon)[2C^f_{00}+(t-m^2_{A_0})C^f_1+m_Z^2C^e_1\nonumber\\
&&\hspace{0.9cm}+(s-m^2_Z-m^2_{A_0})(C^e_1-C^f_2) +
m^2_{A_0}(2C^f_1-C^e_2)-B_0(t,0,0)]+m_Z^2C^i_0\}\nonumber\\
&&\hspace{0.9cm}+\frac{-4gm_b\alpha_sA_3T_{Z}(a,b)R^{\tilde
b}_{a1}R^{\tilde b}_{b1}}{3\pi cos\theta_{w }t}C^q_{00},\nonumber
\\
&& f^c_1=\epsilon\frac{-4im_bA_3\alpha_s}{3\pi
}(C_V+C_A)(C^d_1-C^h_2),\nonumber
\\ && f^c_2=\frac{-4im_bA_3\alpha_s}{3\pi
u}(C_V+C_A)[m^2_{A_0}C^j_0-B_0(u,0,0)]+\frac{4m_{\tilde{g}}\alpha_s}{3\pi
u}\hat{G}^{\tilde b }_3(a,b)R^{\tilde b}_{a1}R^{\tilde
b}_{b2}C^s_0,\nonumber\\
&&\hspace{0.9cm}+\epsilon\frac{-4im_bA_3\alpha_s}{3\pi
u}(C_V+C_A)[uC^d_1+uC^d_2-uC^h_1-uC^h_2+(u-m^2_Z)C^h_1\nonumber\\
&&\hspace{0.9cm}-m_Z^2(C^d_2-2C^h_1)
+(s-m_Z^2-m_{A^0}^2)(C^d_1-C^h_2)+m_{A^0}^2C^d_1+B_0(u,0,0)],\nonumber
\\&& f^c_3=\epsilon\frac{8im_bA_3\alpha_s}{3\pi
}C_V(C^d_1-C^h_2),\nonumber
\\&& f^c_4=\frac{8im_bA_3\alpha_sC_V}{3\pi
u}[m^2_{A_0}C^j_0-B_0(u,0,0)]\nonumber\\
&&\hspace{0.9cm}+\frac{4m_{\tilde{g}}\hat{G}^{\tilde b
}_3(a,b)\alpha_s }{3\pi u}[R^{\tilde b}_{a2}R^{\tilde
b}_{b1}(C_V-C_A)C^s_0-R^{\tilde b}_{a1}R^{\tilde
b}_{b2}(C_V+C_A)C^s_0]\nonumber\\
&&\hspace{0.9cm}+\epsilon\frac{8im_bA_3\alpha_s}{3\pi
u}C_V[uC^d_1+uC^d_2-uC^h_1-uC^h_2+(u-m^2_Z)C^h_1\nonumber\\
&&\hspace{0.9cm}-m_Z^2(C^d_2-2C^h_1)
+(s-m_Z^2-m_{A^0}^2)(C^d_1-C^h_2)+m_{A^0}^2C^d_1+B_0(u,0,0)],\nonumber
\\ && f^c_5=\frac{-2im_bA_3\alpha_s}{3\pi
u}(C_V+C_A)[m^2_{A_0}C^j_0-B_0(u,0,0)]+\frac{4m_{\tilde{g}}\alpha_s}{3\pi
u}\hat{G}^{\tilde b }_3(a,b)R^{\tilde b}_{a1}R^{\tilde
b}_{b2}C^s_0\nonumber\\
&&\hspace{0.9cm}+\epsilon\frac{-2im_bA_3\alpha_s}{3\pi
u}(C_V+C_A)[uC^h_1+(s-m_Z^2-m_{A^0}^2)(C^d_1-C^h_2)\nonumber\\
&&\hspace{0.9cm}-m^2_ZC^d_2+m_{A^0}^2C^d_1+B_0(u,0,0)], \nonumber
\\ && f^c_6=\frac{4im_bA_3\alpha_sC_V}{3\pi
u}[m^2_{A_0}C^j_0-B_0(u,0,0)]\nonumber\\
&&\hspace{0.9cm}+\frac{4m_{\tilde{g}}\hat{G}^{\tilde b
}_3(a,b)\alpha_s }{3\pi u}[R^{\tilde b}_{a2}R^{\tilde
b}_{b1}(C_V-C_A)C^s_0-R^{\tilde b}_{a1}R^{\tilde
b}_{b2}(C_V+C_A)C^s_0]\nonumber\\
&&\hspace{0.9cm}+\epsilon\frac{4im_bA_3\alpha_s}{3\pi
u}C_V[uC^h_1+(s-m_Z^2-m_{A^0}^2)(C^d_1-C^h_2)\nonumber\\
&&\hspace{0.9cm}-m^2_ZC^d_2+m_{A^0}^2C^d_1+B_0(u,0,0)], \nonumber
\\
&& f^d_1=\frac{4i(C_V-C_A)m_bA_3\alpha_s}{3\pi
t}[C^k_0m^2_{A_0}-B_0(t,0,0)]+\frac{-4(C_V-C_A)m_{\tilde g
}\alpha_s\hat{G}^{\tilde b }_3(a,b)R^{\tilde b}_{a1}R^{\tilde
b}_{b2}}{3\pi t}C^t_0\nonumber\\
&&\hspace{0.9cm}+\epsilon\frac{4i(C_V-C_A)m_bA_3\alpha_s}{3\pi
t}[C^b_1m^2_{A_0}+B_0(t,0,0)],\nonumber
\\&& f^d_3=\frac{-8iC_Vm_bA_3\alpha_s}{3\pi
t}[C^k_0m^2_{A_0}-B_0(t,0,0)]
+\frac{-4(C_V+C_A)m_{\tilde g}\alpha_s\hat{G}^{\tilde b
}_3(a,b)R^{\tilde b}_{a2}R^{\tilde b}_{b1}}{3\pi t}C^t_0\nonumber\\
&&\hspace{0.9cm}+\frac{4(C_A-C_V)m_{\tilde
g}\alpha_s\hat{G}^{\tilde b }_3(a,b)R^{\tilde b}_{a1}R^{\tilde
b}_{b2}}{3\pi
t}C^t_0\nonumber\\
&&\hspace{0.9cm}+\epsilon\frac{-8iC_Vm_bA_3\alpha_s}{3\pi
t}[C^b_1m^2_{A_0}+B_0(t,0,0)],\nonumber\\ &&
f^d_5=\frac{-f^d_1}{2},\nonumber
\\&& f^d_6=\frac{-f^d_3}{2},\nonumber
\\
&& f^e_2=\frac{-4i(C_V+C_A)m_bA_3\alpha_s}{3\pi
u}[2C^g_{00}+uC^g_{11}+uC^g_{12}+m^2_Z(C^c_1+C^l_0)]\nonumber\\
&&\hspace{0.9cm}+\frac{4gm_bA_3\alpha_s}{3\pi
ucos\theta_{w}}T_{Z}(a,b)R^{\tilde b}_{a2}R^{\tilde
b}_{b2}(2C^o_{00}-uC^o_{12}),\nonumber\\
&& f^e_4=\frac{8iC_Vm_bA_3\alpha_s}{3\pi
u}[2C^g_{00}+uC^g_{11}+uC^g_{12}+m^2_Z(C^c_1+C^l_0)]\nonumber\\
&&\hspace{0.9cm}+\frac{-4gm_bA_3\alpha_s}{3\pi
ucos\theta_{w}}T_{Z}(a,b)(2C^o_{00}-uC^o_{12})(R^{\tilde
b}_{a1}R^{\tilde
b}_{b1}+R^{\tilde b}_{a2}R^{\tilde b}_{b2}),\nonumber\\
&& f^e_5=\frac{-2i(C_V+C_A)m_bA_3\alpha_s}{3\pi
u}[2C^g_{00}+uC^g_{11}+uC^g_{12}+m^2_Z(C^c_1+C^l_0)]\nonumber\\
&&\hspace{0.9cm}+\frac{4gm_bA_3\alpha_s}{3\pi
ucos\theta_{w}}T_{Z}(a,b)R^{\tilde b}_{a2}R^{\tilde
b}_{b2}C^o_{00},\nonumber
\\&& f^e_6=\frac{4iC_Vm_bA_3\alpha_s}{3\pi
u}[2C^g_{00}+m^2_Z(C^c_1+C^l_0)] +\frac{-4gm_bA_3\alpha_s}{3\pi
ucos\theta_{w}}T_{Z}(a,b)C^o_{00}(R^{\tilde b}_{a1}R^{\tilde
b}_{b1}+R^{\tilde b}_{a2}R^{\tilde b}_{b2}),\nonumber
\\
&& f^f_1=\frac{2i(C_V-C_A)m_bA_3\alpha_s}{3\pi t^2}{R^{\tilde b
}_{a1}}^2[-2m^2_{\tilde g}B_0(0,m^2_{\tilde g},m^2_{\tilde g})
+m^2_{\tilde g}B_0(t,m^2_{\tilde g},m^2_{{\tilde
b}_a})-2m^2_{\tilde g}\nonumber\\
&&\hspace{0.9cm}+ 2m^2_{{\tilde b}_a}B_0(0,m^2_{{\tilde
b}_a},m^2_{{\tilde b}_a})-m^2_{{\tilde b}_a}B_0(t,m^2_{{\tilde
g}},m^2_{{\tilde b}_a}) -2m^2_{{\tilde b}_a}+tB_0(t,m^2_{{\tilde
g}},m^2_{{\tilde b}_a})\nonumber\\
&&\hspace{0.9cm}+ 2m^2_{{\tilde b}_a}B_0(0,m^2_{{\tilde
g}},m^2_{{\tilde b}_a})]+\frac{2i(C_V-C_A)m_bA_3\alpha_s}{3\pi
t^2}(1-\epsilon)tB_0(t,0,0),\nonumber \\
&& f^f_3=\frac{-2im_bA_3\alpha_s}{3\pi t^2}[(C_V-C_A){R^{\tilde b
}_{a1}}^2+(C_V+C_A){R^{\tilde b }_{a2}}^2][-2m^2_{\tilde
g}B_0(0,m^2_{\tilde g},m^2_{\tilde g}) +m^2_{\tilde
g}B_0(t,m^2_{\tilde g},m^2_{{\tilde b}_a})\nonumber\\
&&\hspace{0.9cm}+2m^2_{{\tilde b}_a}B_0(0,m^2_{{\tilde
b}_a},m^2_{{\tilde b}_a})-2m^2_{\tilde g}-m^2_{{\tilde
b}_a}B_0(t,m^2_{{\tilde g}},m^2_{{\tilde b}_a}) -2m^2_{{\tilde
b}_a}+tB_0(t,m^2_{{\tilde g}},m^2_{{\tilde b}_a})
\nonumber\\&&\hspace{0.9cm}+ 2m^2_{{\tilde b}_a}B_0(0,m^2_{{\tilde
g}},m^2_{{\tilde b}_a})] +\frac{-2im_bA_3C_V\alpha_s}{3\pi
t^2}(1-\epsilon)tB_0(t,0,0),\nonumber
\\
&& f^g_2=\frac{-2i(C_V+C_A)m_bA_3\alpha_s}{3\pi t^2}{R^{\tilde b
}_{a2}}^2[-2m^2_{\tilde g}B_0(0,m^2_{\tilde g},m^2_{\tilde g})
+m^2_{\tilde g}B_0(t,m^2_{\tilde g},m^2_{{\tilde b}_a})+
2m^2_{{\tilde
b}_a}B_0(0,m^2_{{\tilde b}_a},m^2_{{\tilde b}_a})\nonumber\\
&&\hspace{0.9cm}-2m^2_{\tilde g}-m^2_{{\tilde
b}_a}B_0(u,m^2_{{\tilde g}},m^2_{{\tilde b}_a}) -2m^2_{{\tilde
b}_a}+uB_0(u,m^2_{{\tilde g}},m^2_{{\tilde b}_a})+
2m^2_{{\tilde b}_a}B_0(0,m^2_{{\tilde g}},m^2_{{\tilde b}_a})] \nonumber\\
&&\hspace{0.9cm}+\frac{-2i(C_V+C_A)m_bA_3\alpha_s}{3\pi
t^2}(1-\epsilon)uB_0(u,0,0),\nonumber \\
&& f^g_4=\frac{2im_bA_3\alpha_s}{3\pi u^2}[(C_V-C_A){R^{\tilde b
}_{a1}}^2+(C_V+C_A){R^{\tilde b }_{a2}}^2][-2m^2_{\tilde
g}B_0(0,m^2_{\tilde g},m^2_{\tilde g}) +m^2_{\tilde
g}B_0(u,m^2_{\tilde g},m^2_{{\tilde b}_a})\nonumber\\
&&\hspace{0.9cm}+ 2m^2_{{\tilde b}_a}B_0(0,m^2_{{\tilde
b}_a},m^2_{{\tilde b}_a})-2m^2_{\tilde g}-m^2_{{\tilde
b}_a}B_0(u,m^2_{{\tilde g}},m^2_{{\tilde b}_a}) -2m^2_{{\tilde
b}_a}+uB_0(u,m^2_{{\tilde g}},m^2_{{\tilde
b}_a})\nonumber\\&&\hspace{0.9cm}+ 2m^2_{{\tilde
b}_a}B_0(0,m^2_{{\tilde g}},m^2_{{\tilde b}_a})]
+\frac{2im_bA_3C_V\alpha_s}{3\pi
u^2}(1-\epsilon)uB_0(u,0,0),\nonumber \\
&& f^g_5=\frac{-f^g_4}{2},\nonumber \\
&& f^f_6=\frac{-f^g_2}{2}.\nonumber
\end{eqnarray}
 For the box diagrams(a)-(d) in Fig.~3, we find, respectively,
 \begin{eqnarray}
&&f^{Box(a)}_1=\frac{-4im_bA_3(C_V-C_A)\alpha_s}{3\pi}[-C^e_2-sD^a_0
+uD^b_2-(t-m^2_{A^0})(D^b_2+D^b_3-D^a_1-D^a_3)\nonumber\\
&&\hspace{1.6cm}+(1+\epsilon)(C^y_0-C^x_0)]\nonumber\\
&&\hspace{1.6cm}+\epsilon\frac{-4im_bA_3(C_V-C_A)\alpha_s}{3\pi}[
C^e_2+C^u_1+(u-m^2_{A^0})(D^a_{11}+D^a_{13})-(t-m^2_{A^0})(D^b_2+D^b_3)\nonumber\\
&&\hspace{1.6cm}
-m^2_{A^0}(D^a_1+D^a_{13}+D^a_3+D^a_{33})],\nonumber\\
&&f^{Box(a)}_2=\frac{4im_bA_3(C_V-C_A)\alpha_s}{3\pi}
[m^2_{A^0}(D^a_0+D^b_2)-(t-m^2_{A^0})(D^a_1+D^a_2+D^a_3)-(u-m^2_{A^0})D^b_3\nonumber\\
&&\hspace{1.6cm}-C^a_0+C^z_0-C^x_0]
\nonumber\\
&&\hspace{1.6cm}+\epsilon\frac{4im_bA_3(C_V-C_A)\alpha_s}{3\pi} [
2D^a_{00}-(u-m^2_{A^0})(D^a_{11}+D^a_{12}+D^a_{13}+D^b_3)\nonumber\\
&&\hspace{1.6cm}+m^2_{A^0}
(D^a_{13}+D^a_{23}+D^a_{33}+D^b_2)],\nonumber\\ &&
f^{Box(a)}_3=\frac{-2C_Vf^{Box(a)}_1}{C_V-C_A},\nonumber\\ &&
f^{Box(a)}_4=\frac{-2C_Vf^{Box(a)}_2}{C_V-C_A},\nonumber\\ &&
f^{Box(a)}_5=\frac{2im_bA_3(C_V-C_A)\alpha_s}{3\pi}[
-sD^a_0+(u-t)D^a_0+C^y_0+C_0^z-2C_0^x]\nonumber\\
&&\hspace{1.6cm}+\epsilon\frac{2im_bA_3(C_V-C_A)\alpha_s}{3\pi}[
2D^a_{00}-(t-m^2_{A^0})D^b_2-m^2_{A^0}D^a_3-C^x_0],\nonumber\\ &&
f^{Box(a)}_6=\frac{-2C_Vf^{Box(a)}_5}{C_V-C_A},\nonumber
\\
&&f^{Box(b)}_1=\frac{4igm_{\tilde g}\alpha_s}{3\pi cos\theta_{w}
cos\theta_{w} }T_Z(l,a){\hat G}_3^{\tilde b}(l,b)R^{\tilde
b}_{a1}R^{\tilde b}_{b2}(D^e_1+D^e_3+D^f_0),\nonumber\\
&&f^{Box(b)}_2=\frac{4igm_{\tilde g}\alpha_s}{3\pi cos\theta_{w}
cos\theta_{w} }T_Z(l,a){\hat G}_3^{\tilde b}(l,b)R^{\tilde
b}_{a1}R^{\tilde b}_{b2}(D^e_1+D^e_2+D^e_3+D^f_0),\nonumber\\
&&f^{Box(b)}_3=\frac{4igm_{\tilde g}\alpha_s}{3\pi cos\theta_{w}
cos\theta_{w} }T_Z(l,a){\hat G}_3^{\tilde b}(l,b)(R^{\tilde
b}_{b1}-R^{\tilde b}_{a1}R^{\tilde
b}_{b2})(D^e_1+D^e_3+D^f_0),\nonumber\\
&&f^{Box(b)}_4=\frac{4igm_{\tilde g}\alpha_s}{3\pi cos\theta_{w}
cos\theta_{w} }T_Z(l,a){\hat G}_3^{\tilde b}(l,b)(R^{\tilde
b}_{b1}-R^{\tilde b}_{a1}R^{\tilde
b}_{b2})(D^e_1+D^e_2+D^e_3+D^f_0),\nonumber
\\
&&f^{Box(c)}_1=\frac{-4im_bA_3(C_V+C_A)\alpha_s}{3\pi}[-sD^c_1
+(u-m^2_Z)(D^d_3-D^c_1)-m^2_ZD^d_2-C^v_0]\nonumber\\
&&\hspace{1.6cm}+\epsilon\frac{-4im_bA_3(C_V+C_A)\alpha_s}{3\pi}[
C^d_1-C^u_1-C^x_1-(t-m^2_Z)D^c_{11}+(u-m^2_Z)D^d_3\nonumber\\
&&\hspace{1.6cm}+m^2_Z(D^c_1+D^c_{13})],\nonumber\\
&&f^{Box(c)}_2=\frac{-4im_bA_3(C_V+C_A)\alpha_s}{3\pi}[C^v_1-sD^c_1-sD^c_2
-(t-m^2_Z)D^d_3-(u-m^2_Z)(D^c_1+D^c_2)\nonumber\\
&&\hspace{1.6cm}+m^2_Z(D^c_0+D^d_2)-C^a_0]\nonumber\\
&&\hspace{1.6cm}+\epsilon\frac{-4im_bA_3(C_V+C_A)\alpha_s}{3\pi}[C^d_1+C^d_2-C^x_1
-(t-m^2_Z)(D^c_{11}+D^c_{12}+D^d_{3})\nonumber\\
&&\hspace{1.6cm}+m^2_Z(D^c_{13}+D^c_{23}+D^d_2)+C^v_0],\nonumber
\\ &&
f^{Box(c)}_3=\frac{-2C_Vf^{Box(b)}_1}{C_V+C_A},\nonumber\\ &&
f^{Box(c)}_4=\frac{-2C_Vf^{Box(b)}_2}{C_V+C_A},\nonumber\\ &&
f^{Box(c)}_5=\frac{2im_bA_3(C_V+C_A)\alpha_s}{3\pi}[-sD^c_0+(t-u)D^d_2+C^l_0+C^v_0-2C^x_0]\nonumber\\
&&\hspace{1.6cm}+\epsilon\frac{2im_bA_3(C_V+C_A)\alpha_s}{3\pi}[2D^c_{00}-(u-m^2_{A^0})D^d_2-m^2_ZD^c_{3}-C^x_0],\nonumber
\\ &&
f^{Box(c)}_6=\frac{-2C_Vf^{Box(b)}_5}{C_V+C_A},\nonumber\\
&&f^{Box(d)}_1=\frac{4igm_{\tilde g}\alpha_s}{3\pi cos\theta_{w}
cos\theta_{w} }T_Z(l,a){\hat G}_3^{\tilde b}(l,b)R^{\tilde
b}_{a1}R^{\tilde b}_{b2}D^h_1,\nonumber\\
&&f^{Box(d)}_2=\frac{4igm_{\tilde g}\alpha_s}{3\pi cos\theta_{w}
cos\theta_{w} }T_Z(l,a){\hat G}_3^{\tilde b}(l,b)R^{\tilde
b}_{a1}R^{\tilde b}_{b2}(D^h_1+D^h_2),\nonumber\\
&&f^{Box(d)}_3=\frac{4igm_{\tilde g}\alpha_s}{3\pi cos\theta_{w}
cos\theta_{w} }T_Z(l,a){\hat G}_3^{\tilde b}(l,b)(R^{\tilde
b}_{b1}-R^{\tilde b}_{a1}R^{\tilde
b}_{b2})D^h_1,\nonumber\\
&&f^{Box(b)}_4=\frac{4igm_{\tilde g}\alpha_s}{3\pi cos\theta_{w}
}T_Z(l,a){\hat G}_3^{\tilde b}(l,b)(R^{\tilde b}_{a2}R^{\tilde
b}_{b1}-R^{\tilde b}_{a1}R^{\tilde b}_{b2})(D^h_1+D^h_2).\nonumber
\end{eqnarray}

\newpage

\begin{figure}[h!]
\vspace{1.0cm} \centerline{\epsfig{file=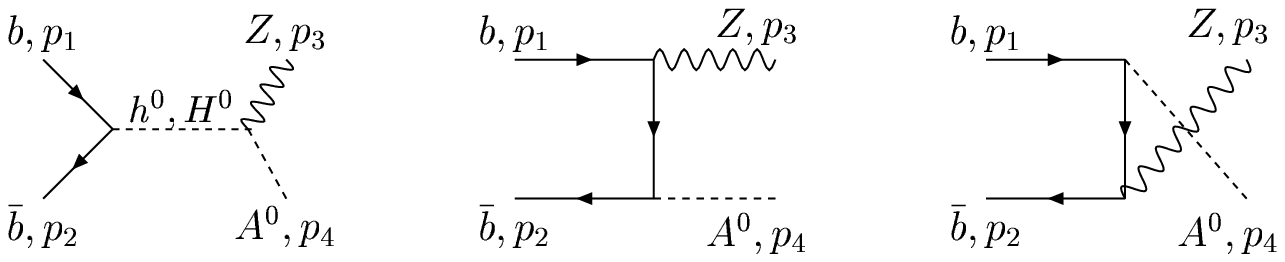,
width=400pt}} \caption[]{Leading order Feynman diagrams for
$b\bar{b}\rightarrow A^0Z^0$.}
\end{figure}
\begin{figure}[h!]
\vspace{1.0cm} \centerline{\epsfig{file=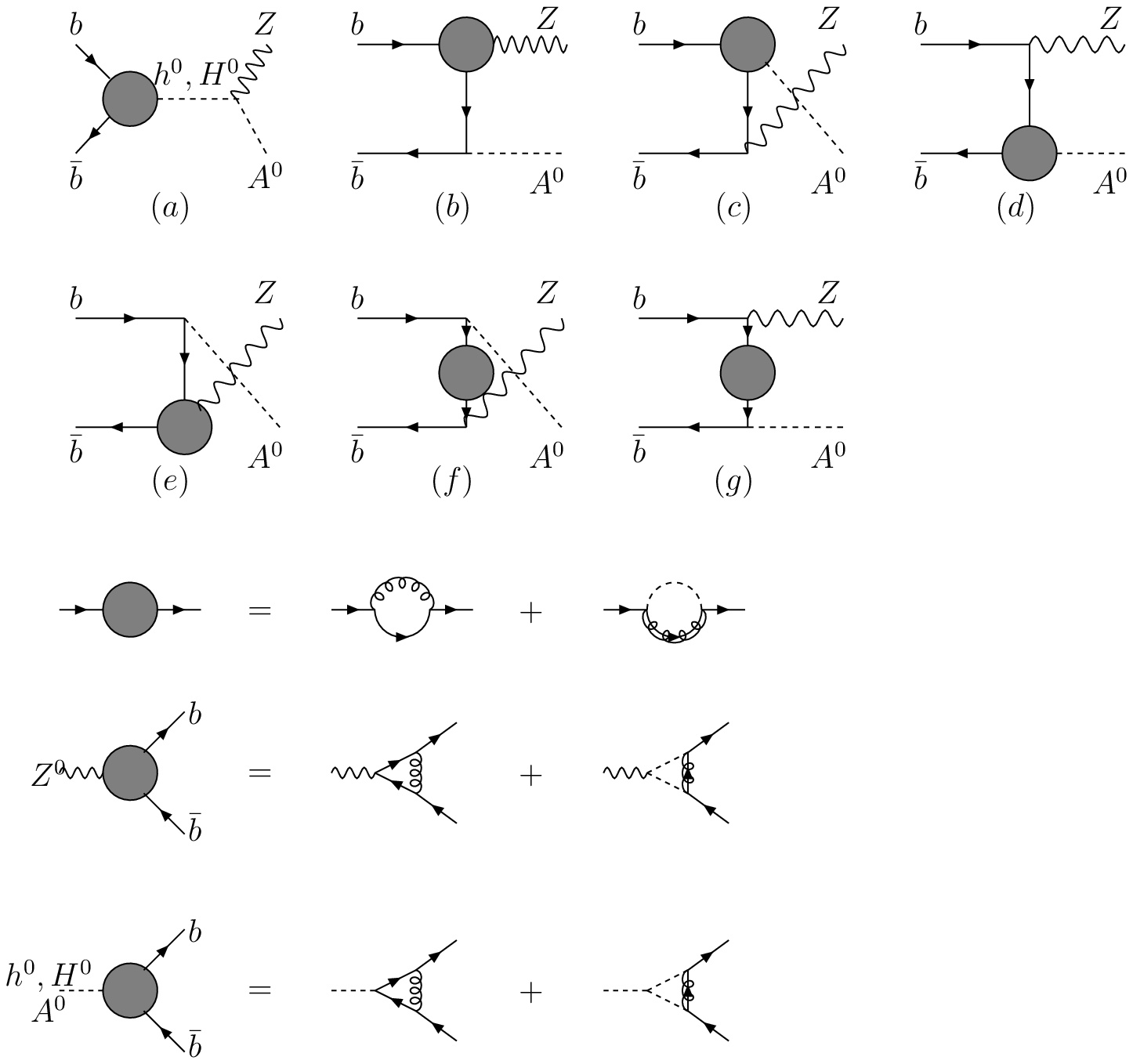,
width=380pt}} \caption[]{One-loop virtual diagrams,
including self-energy and vertex corrections for\\
$b\bar{b}\rightarrow A^0Z^0$.}
\end{figure}
\newpage
\begin{figure}[h!]
\vspace{1.0cm} \centerline{\epsfig{file=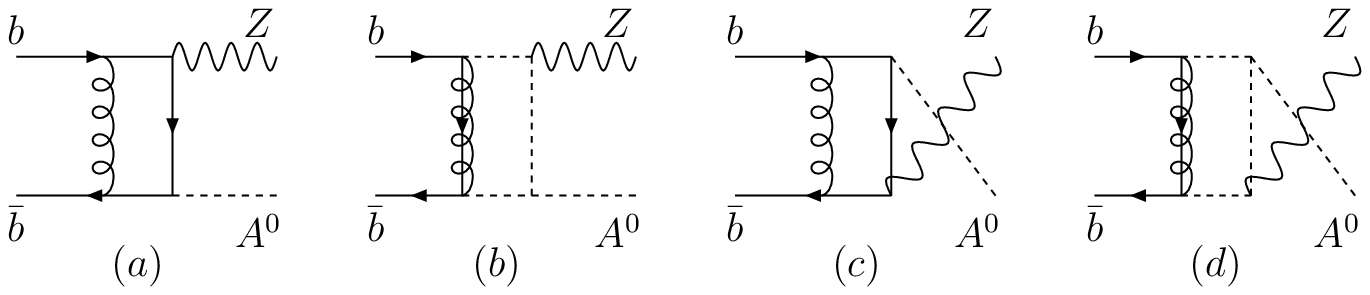,
width=380pt}} \caption[]{Box diagrams
 for
$b\bar{b}\rightarrow A^0Z^0$.}
\end{figure}
\begin{figure}[h!]
\vspace{1.0cm} \centerline{\epsfig{file=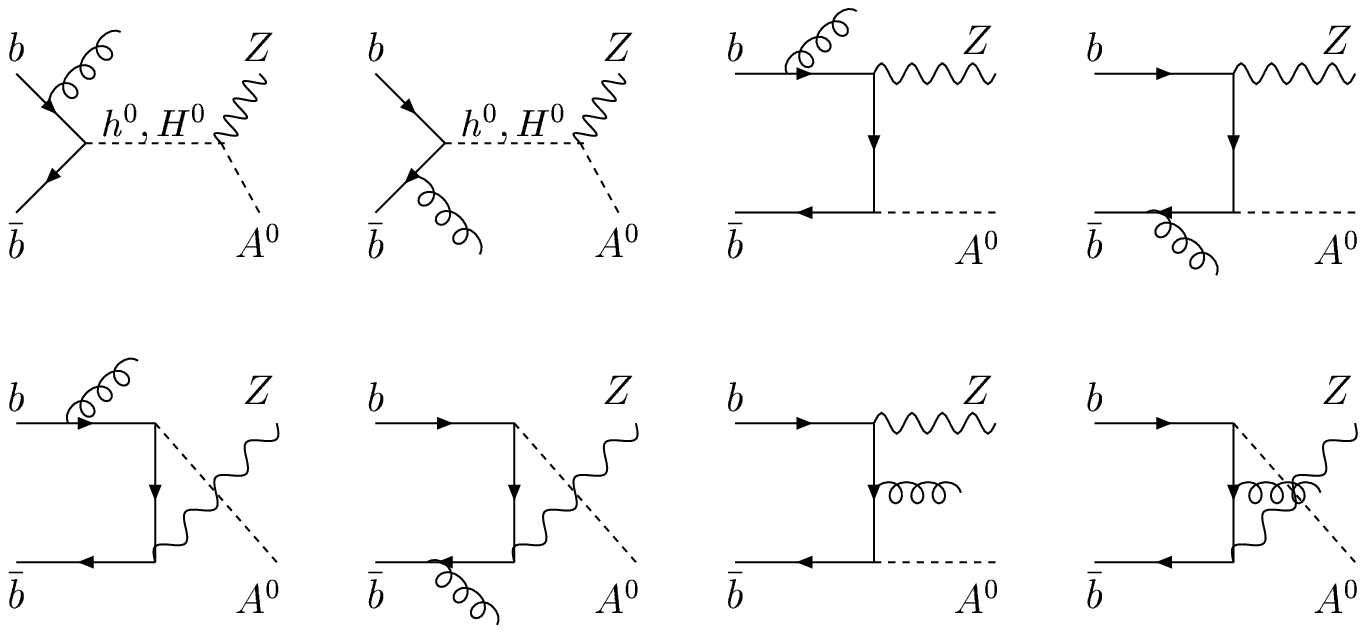,
width=380pt}} \caption[]{Feynman diagrams
 for the real gluon emission contributions.}
\end{figure}
\begin{figure}[h!]
\vspace{1.0cm} \centerline{\epsfig{file=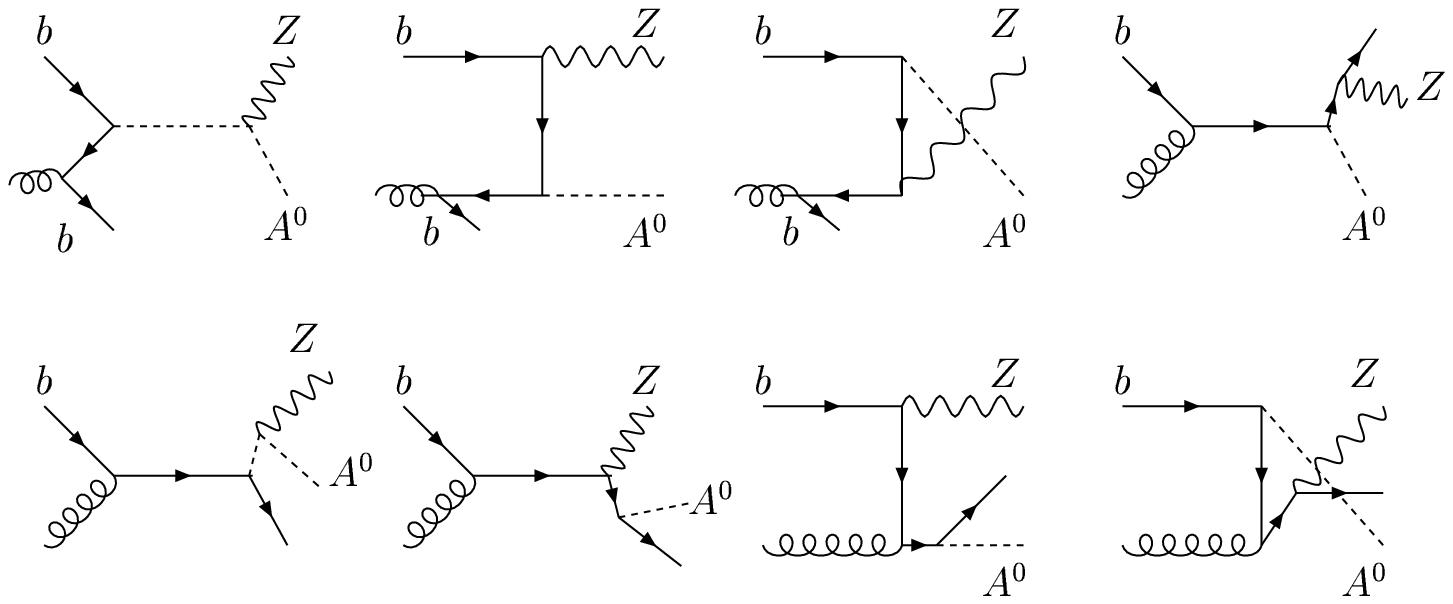,
width=380pt}} \caption[]{Feynman diagrams
 for the emission of a massless bottom
quark contributions.}
\end{figure}
\begin{figure}[h!]
\centerline{\epsfig{file=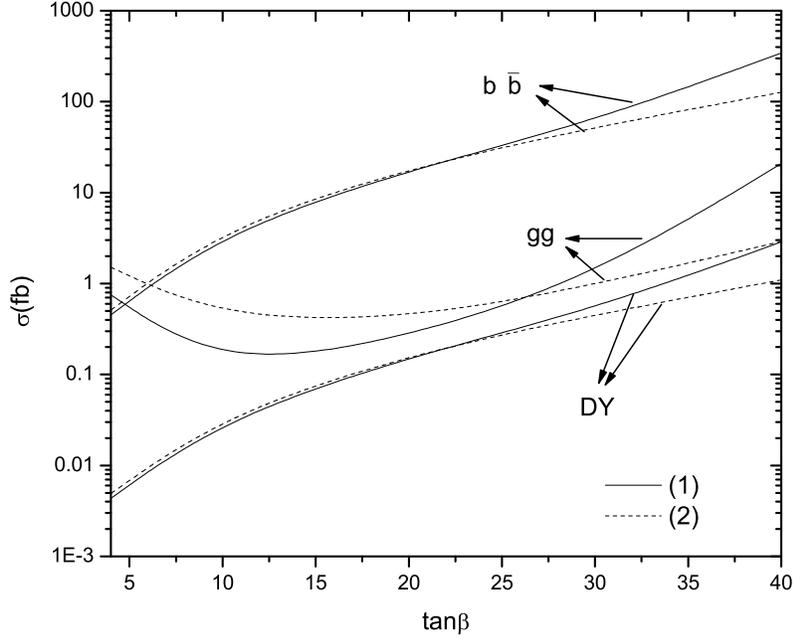,width=350pt}} \caption[]{ LO total
cross sections of $pp\rightarrow A^0Z^0$ via $b\bar{b}$
annihilation, compared with the ones from gluon fusion and Drell-Yan
processes at the LHC, as a function of $\tan\beta$
with $m_b(m_b)=4.25$\,GeV, assuming: (1)
$m_0=200$\,GeV, $m_{\frac{1}{2}}=160$\,GeV, $A_0=100$\,GeV
and $ \mu<0$; (2)
$m_0=150$\,GeV, $m_{\frac{1}{2}}=180$\,GeV, $A_0=300$\,GeV and $ \mu>0$.
\label{ds}}
\end{figure}
\begin{figure}[h!]
\centerline{\epsfig{file=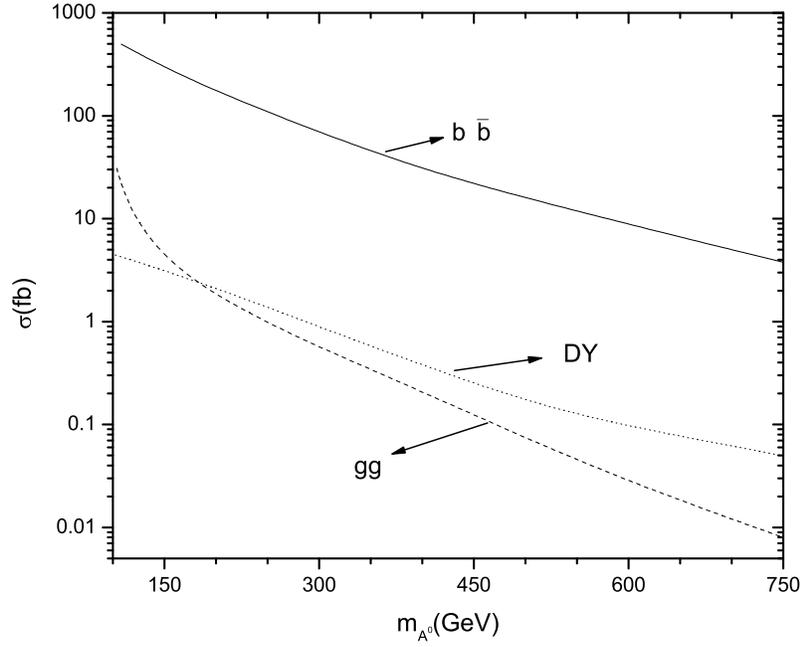,width=350pt}} \caption[]{ LO total
cross sections of $pp\rightarrow A^0Z^0$ via $b\bar{b}$
annihilation, compared with the ones from gluon fusion and Drell-Yan
process at the LHC, as a function of $m_{A^0}$ with $m_b(m_b)=4.25$\,GeV,
 assuming:
$m_{\frac{1}{2}}=160$\,GeV, $A_0=100$\,GeV, $\tan\beta=40$ and $\mu<0$.
\label{ds}}
\end{figure}
\begin{figure}[h!]
\centerline{\epsfig{file=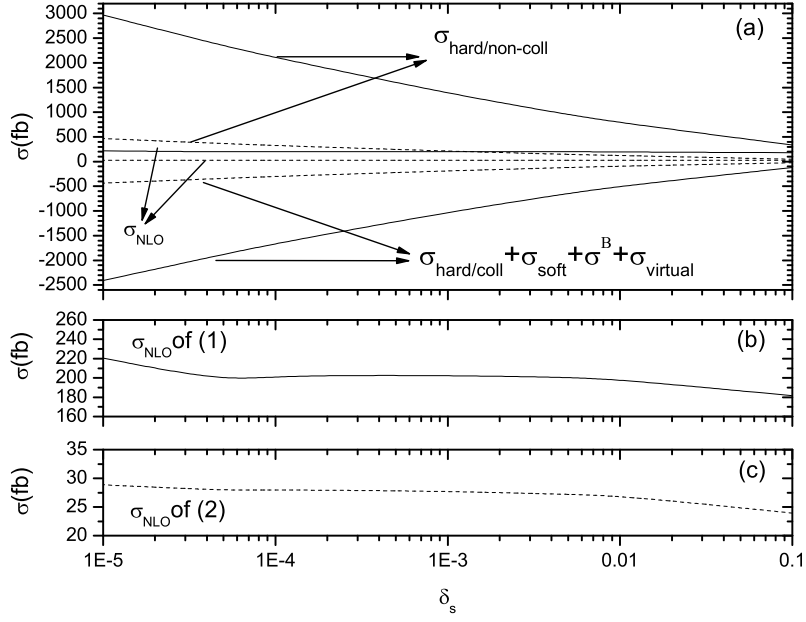,width=350pt}}
\caption[]{Dependence of the NLO total cross sections for the $A^0Z^0$
production at the LHC on the theoretical cutoff cale
$\delta_s$ with $\delta_c=\delta_s/50$, assuming: (1)
$m_0=200$\,GeV, $m_{\frac{1}{2}}=160$\,GeV,
$A_0=100$\,GeV, $\tan\beta=40 $ and $
\mu<0$; (2)
$m_0=150$\,GeV, $m_{\frac{1}{2}}=180$\,GeV, $A_0=300$\,GeV,
$\tan\beta=40 $ and $
\mu>0$.
Here, we take $m_b(m_b)=4.25$\,GeV. In (a), the solid and dotted curves
are the results for model (1) and (2), respectively.
\label{ds}}
\end{figure}
\begin{figure}[h!]
\centerline{\epsfig{file=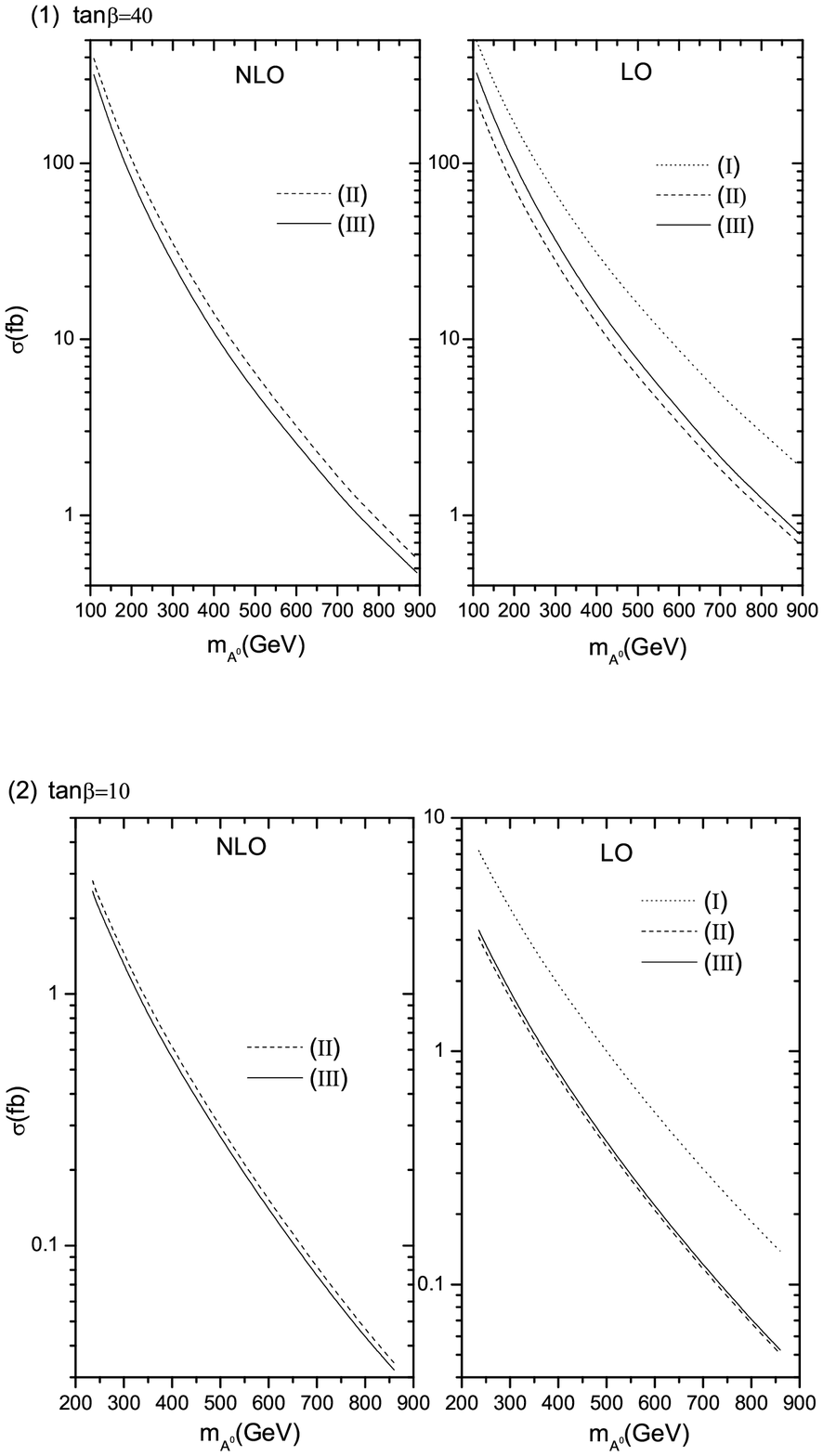,width=300pt}}
\caption[]{Dependence of the total cross section of the $A^0Z^0$
production at the LHC on $m_{A^0}$, assuming
$m_{\frac{1}{2}}=160$\,GeV, $A_0=100$\,GeV, and $\mu<0$ for
$\tan\beta=40$ in Fig.~9(1) and $\tan\beta=10$ in Fig.~9(2).
Three differet calculations were done by using: (I)
$\overline{\rm MS}$ bottom quark mass at the scale $m_b$,
(II) QCD improved bottom quark running mass
at the scale $m_{A^0}$, and
(III) QCD plus SUSY improved bottom quark running mass
at the scale $m_{A^0}$, respectively, to evaluate the
bottom quark Yukawa coupling.
\label{mA0}}
\end{figure}
\begin{figure}[h!]
\centerline{\epsfig{file=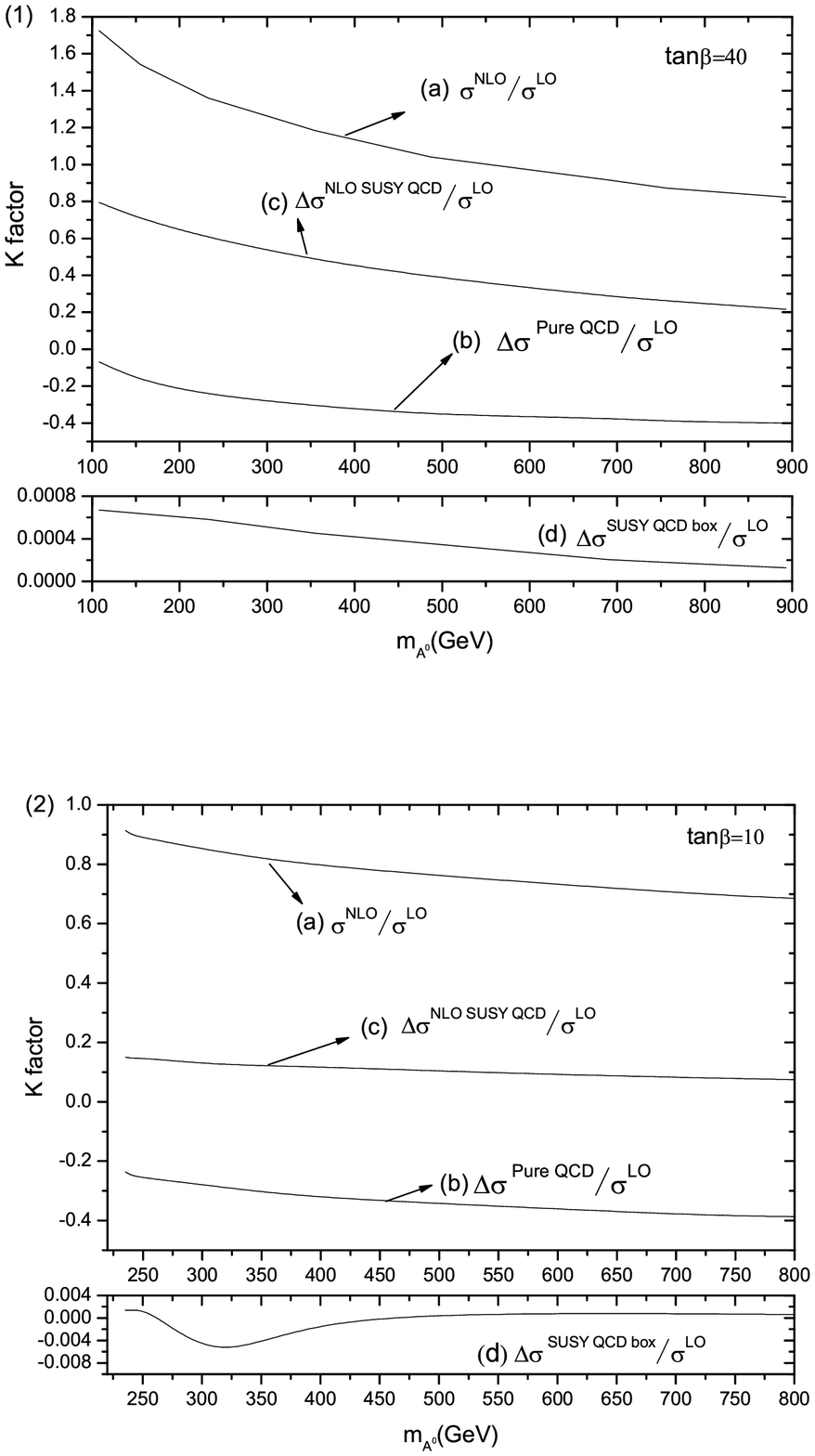,width=300pt}}
 \caption[]{$K$-factor, defined as
$\sigma_{NLO}/\sigma_{LO}$,
 for the $A^0Z^0$ production at
the LHC as a function of $m_{A^0}$, using the QCD improved running
$m_b$ to evaluate the bottom quark Yukawa coupling,
assuming $m_{\frac{1}{2}}=160$\,GeV, $A_0=100$\,GeV, and
$\mu<0$ for $\tan\beta=40$ in Fig.~10(1) and $\tan\beta=10$ in
Fig.~10(2). The full $K$-factor is shown as curve (a), which includes
the pure QCD corrections, shown as curve (b), and SUSY QCD corrections,
shown as curve (c). The contribution from the
SUSY QCD box diagrams is also separately shown as curve (d)
for comparison.
\label{mA0ratioqcd}}
\end{figure}
\begin{figure}[h!]
\centerline{\epsfig{file=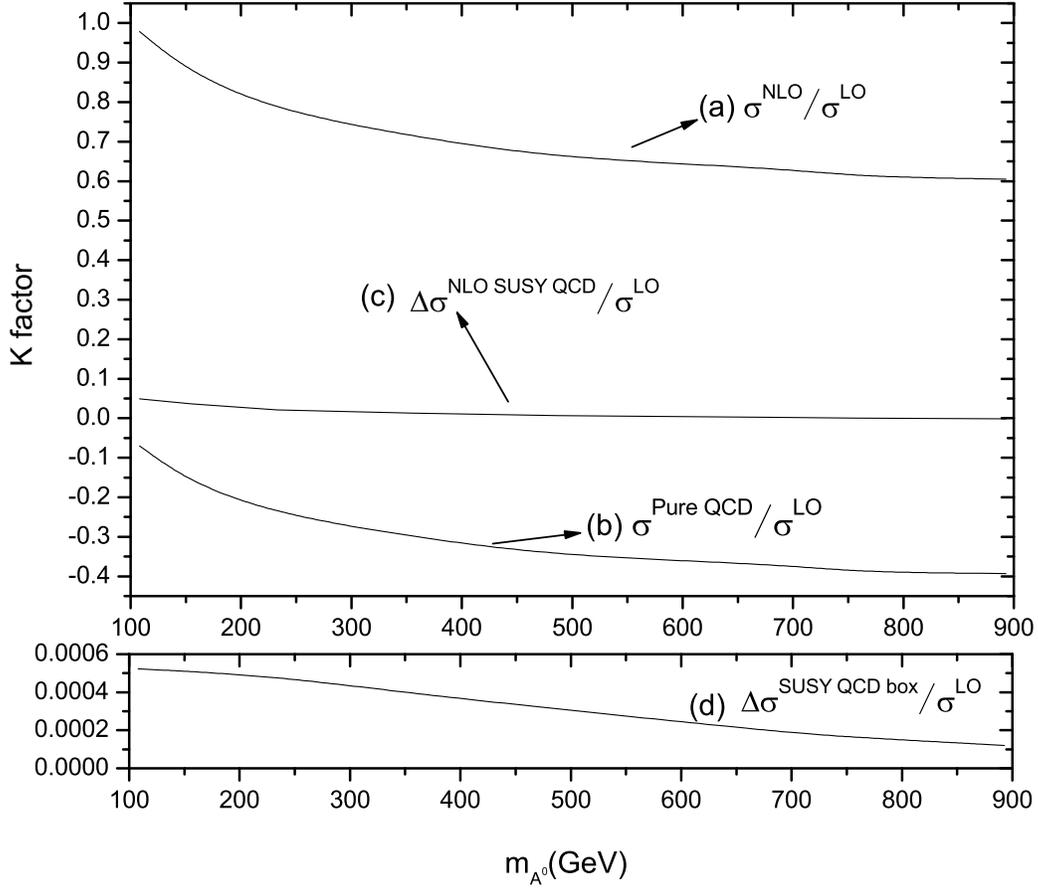,width=450pt}}
 \caption[]{$K$-factor, defined as
$\sigma_{NLO}/\sigma_{LO}$,
 for the $A^0Z^0$ production at
the LHC as a function of $m_{A^0}$, using the
QCD plus SUSY improved bottom quark Yukawa coupling,
assuming $m_{\frac{1}{2}}=160$\,GeV, $A_0=100$\,GeV,
$\mu<0$ and $\tan\beta=40$.
 The full $K$-factor is shown as curve (a), which includes
the pure QCD corrections, shown as curve (b), and SUSY QCD corrections,
shown as curve (c). The contribution from the
SUSY QCD box diagrams is also separately shown as curve (d)
for comparison.
 \label{mA01kfactor}}
\end{figure}
\begin{figure}[h!]
\centerline{\epsfig{file=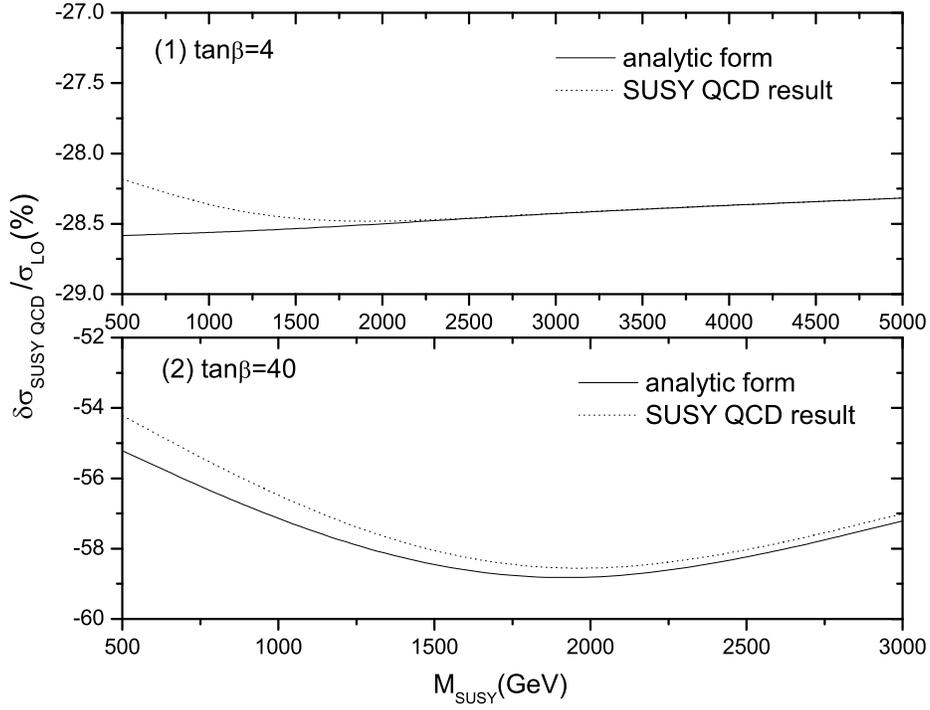,width=400pt}} \caption[
]{Comparison of the SUSY QCD corrections, denoted as
$\delta\sigma_{SUSY QCD}/\sigma_{LO}$, for the $A^0Z^0$ production
at the LHC. The results of using the complete numerical
calculation (dashed curves) and the approximate analytical forms
(solid curves) in the heacvy mass limit are separately shown as a
function of $M_{SUSY}$ with $\tan\beta=4$ and $40$, respectively,
assuming $m_{A^0}=150$\,GeV and $M_{\tilde{Q}}$ $=$
$M_{\tilde{D}}$ $=$ $\mu$ $=$ $A_b$ $=$ $M_{\tilde{g}}$ $\equiv
M_{SUSY}$. Here, the LO cross section is calculated by using
the $\overline{\rm MS}$ bottom quark mass. }
\end{figure}

\begin{figure}[h!]
\centerline{\epsfig{file=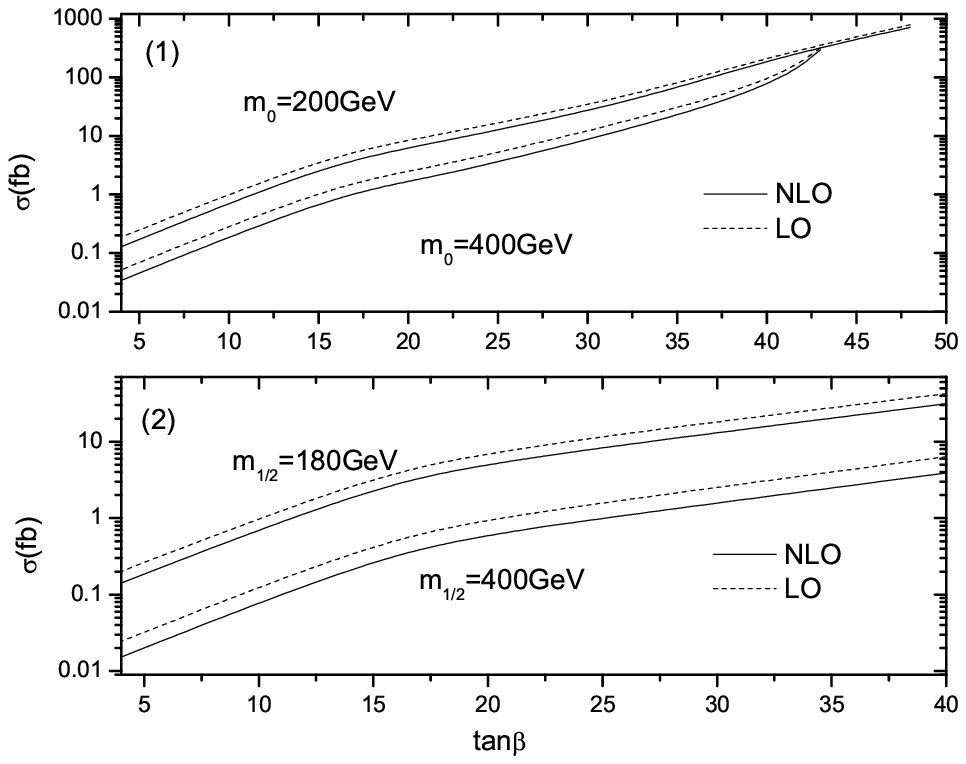,width=400pt}}
\caption[]{Dependence of the total cross sections for the $A^0Z^0$
production at the LHC on $\tan\beta$, assuming:
(1)$m_0=200$\,GeV
and $400$\,GeV, respectively,
$m_{\frac{1}{2}}=160$\,GeV, $A_0=100$\,GeV, and $ \mu<0$;
(2)$m_0=150$\,GeV, $m_{\frac{1}{2}}=180$\,GeV and $400$\,GeV,
respectively,
$A_0=300$\,GeV, and $ \mu>0$. \label{tanb}}
\end{figure}
\begin{figure}[h!]
\centerline{\epsfig{file=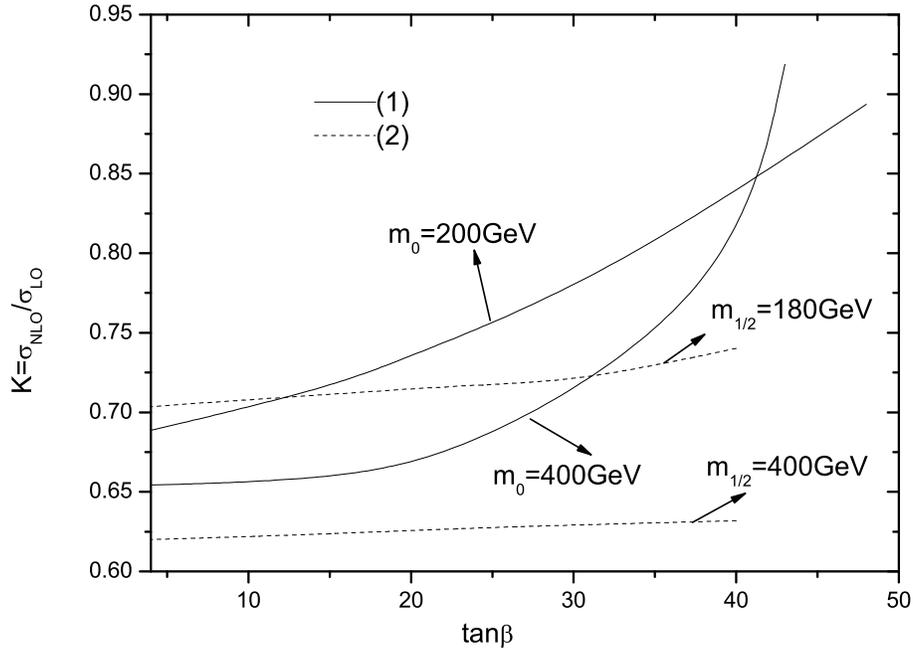,width=400pt}}
\caption[]{Dependence of the $K$-factor, defined as
$\sigma_{NLO}/\sigma_{LO}$, on $\tan\beta$
for the $A^0Z^0$ production at the LHC, assuming:
(1)$m_0=200$\,GeV
and $400$\,GeV, respectively,
$m_{\frac{1}{2}}=160$\,GeV, $A_0=100$\,GeV, and $ \mu<0$;
(2)$m_0=150$\,GeV, $m_{\frac{1}{2}}=180$\,GeV and $400$\,GeV,
respectively,
$A_0=300$\,GeV, and $ \mu>0$.
\label{tanbkfactor}}
\end{figure}
\begin{figure}[h!]
\centerline{\epsfig{file=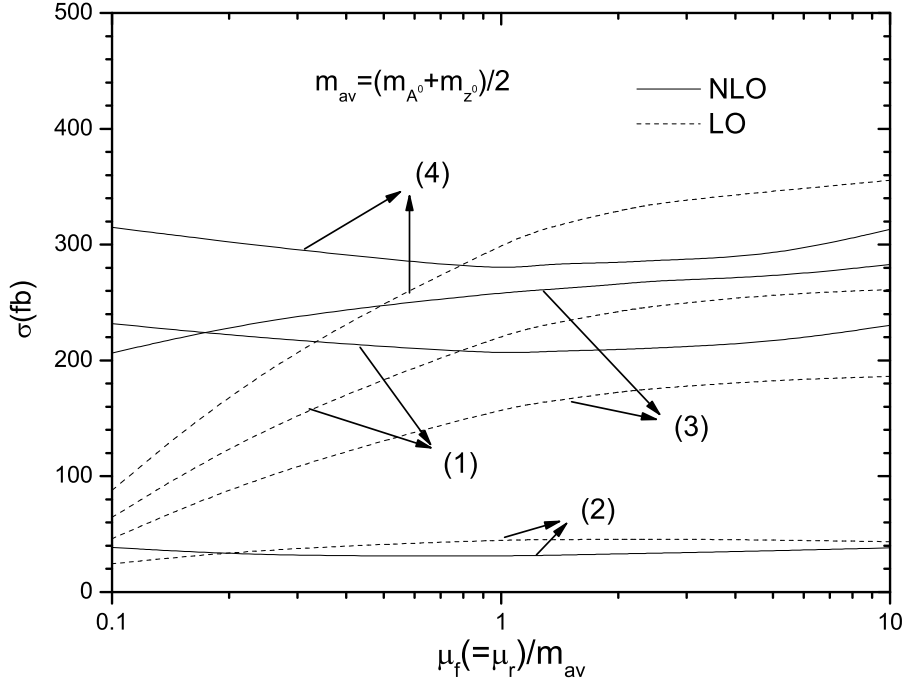,width=400pt}}
\caption[]{Dependence of the total cross sections on the
renormalization/factorization scale $(\mu_r=\mu_f)$ for the
$A^0Z^0$ production at the LHC, assuming: (1) $m_0=200$\,GeV,
$m_{\frac{1}{2}}=160$\,GeV, $A_0=100$\,GeV, $\tan\beta=40$ and $
\mu<0$; (2) $m_0=150$\,GeV, $m_{\frac{1}{2}}=180$\,GeV,
$A_0=300$\,GeV, $\tan\beta=40$ and $ \mu>0$. Here, the QCD plus
SUSY improved bottom quark Yukawa coupling is used. The case of
the curves (3) is similar to (1), but in (3) the pure QCD running
bottom quark mass is used instead. The case of the curves (4) is
similar to (1), but in (4) the contribution from the SUSY-EW
correction in the running bottom quark Yukawa coupling
 is not included, namely,
only the pure QCD and SUYSY-QCD corrections are included.}
 \label{scale}
\end{figure}
\begin{figure}[h!]
\centerline{\epsfig{file=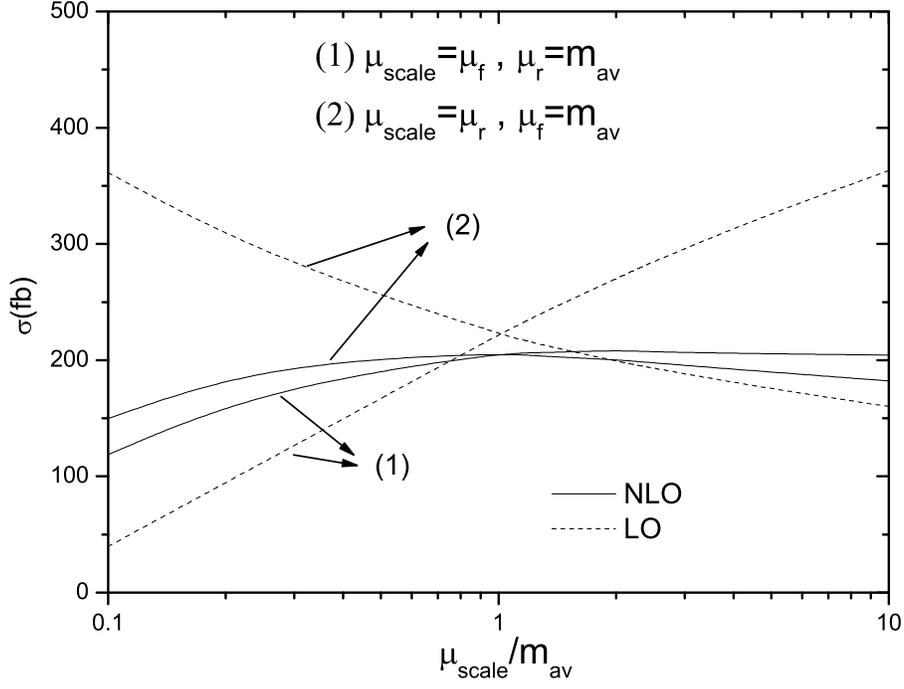,width=400pt}}
\caption[]{Dependence of the total cross sections on the
factorization scale $(\mu_f)$, labelled as case (1), or
renormalization scale $(\mu_r)$, labelled as case (2),
for the $A^0Z^0$ production at the LHC, assuming: $m_0=200$\,GeV,
$m_{\frac{1}{2}}=160$\,GeV, $A_0=100$\,GeV, $\tan\beta=40$ and $
\mu<0$. Here, the QCD plus SUSY improved bottom quark Yukawa
coupling is used and $m_{av}=( m_{A^0} + m_{Z^0} )/2 $.}
 \label{scale2}
\end{figure}
\begin{figure}[h!]
\centerline{\epsfig{file=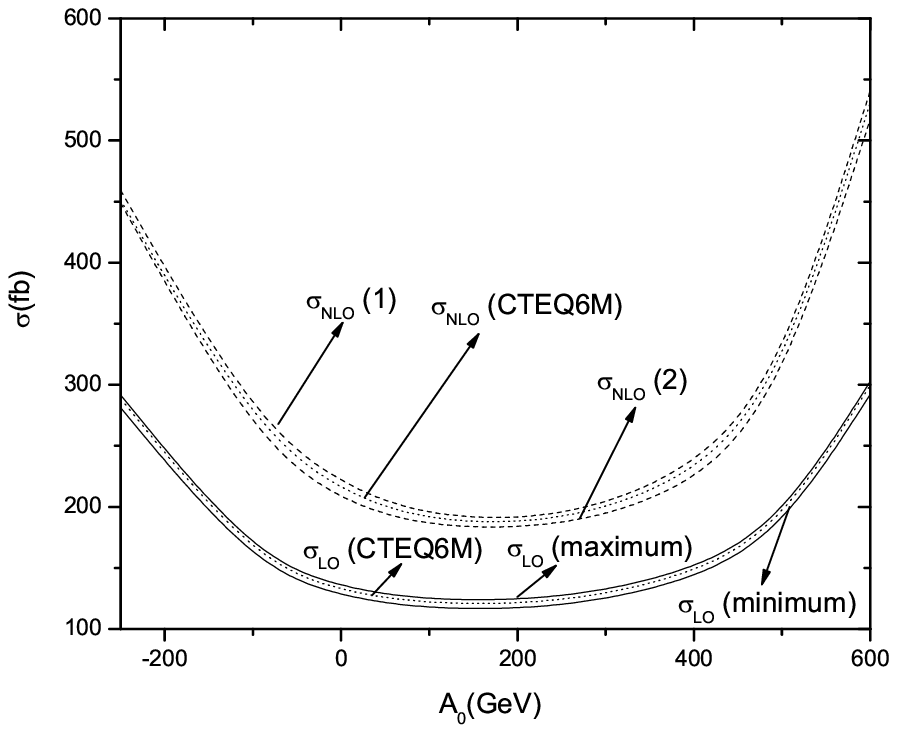,width=400pt}}
 \caption[]{The PDF dependence of the total cross sections for
$pp\rightarrow A^0Z^0$ production at the LHC, as a function of $A_0$, assuming
$m_0=250$\,GeV, $m_{\frac{1}{2}}=160$\,GeV, $\tan\beta=40$ and $
\mu<0$. Here, the QCD running bottom quark mass is used to
evaluate the bottom quark Yukawa coupling.} \label{cteq}
\end{figure}

\begin{figure}[h!]
\centerline{\epsfig{file=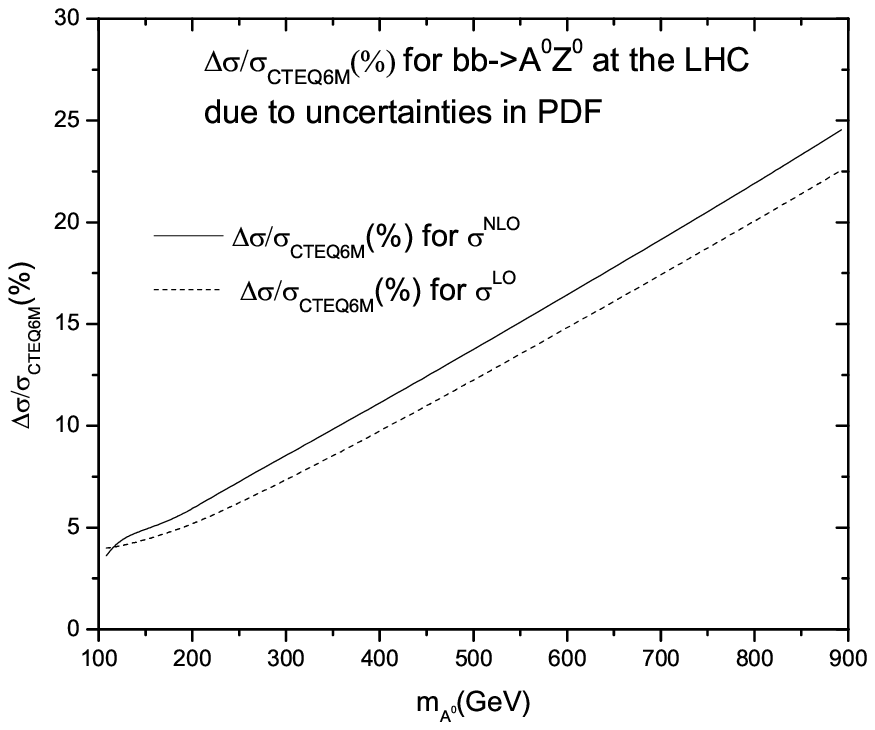,width=400pt}}
 \caption[]{The PDF dependence of the total cross sections for
$pp\rightarrow A^0Z^0$ at the LHC as a function of $m_{A^0}$,
assuming $A_0=100$\,GeV, $m_{\frac{1}{2}}=160$\,GeV,
$\tan\beta=40$ and $ \mu<0$. Here, the QCD running bottom quark
mass is used to evaluate the bottom quark Yukawa coupling.}
\label{PDFU}
\end{figure}
\begin{figure}[h!]
\centerline{\epsfig{file=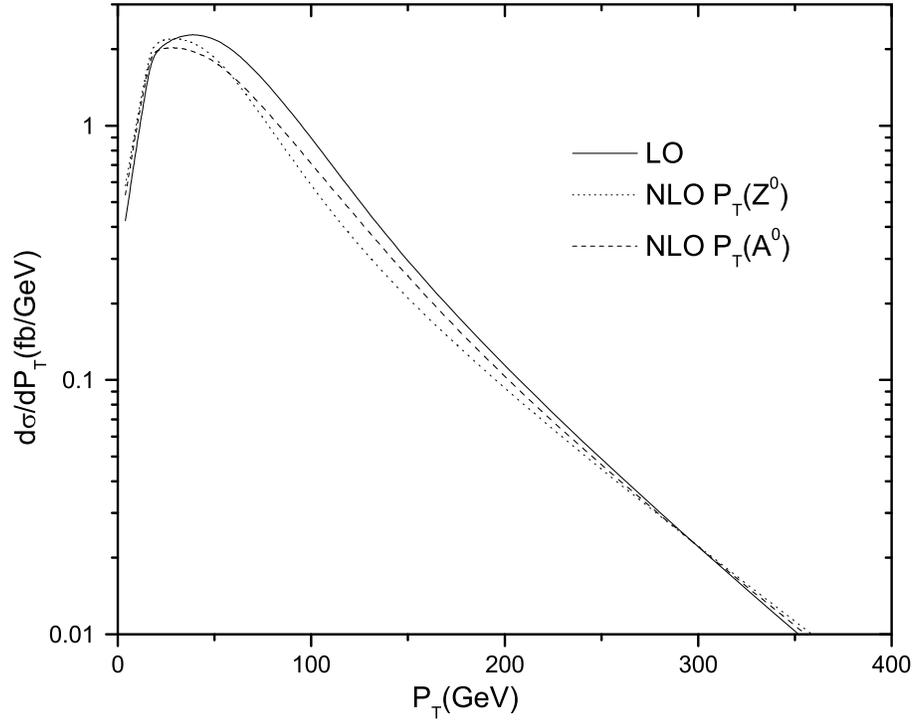,width=400pt}}
\caption[]{Differential cross sections in the transverse momentum
 ($p_T$) of $Z^0$ and $A^0$ bosons,
  for the $A^0Z^0$ production at the
LHC, assuming: $m_0=200$\,GeV, $m_{\frac{1}{2}}=160$\,GeV,
$A_0=100$\,GeV, $\tan\beta=40 $ and $\mu<0$.} \label{pt1}
\end{figure}
\begin{figure}[h!]
\centerline{\epsfig{file=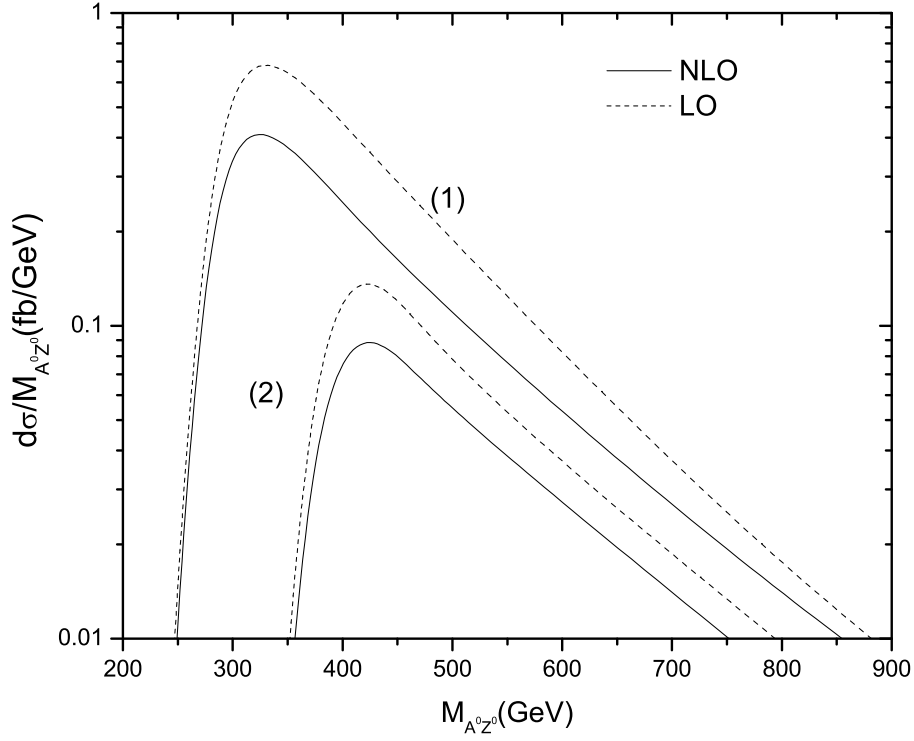,width=400pt}}
\caption[]{Differential cross sections in the invariant mass
 ($M_{A^0Z^0}$), for the $A^0Z^0$ production at the
LHC, assuming: (1) $m_0=200$\,GeV, $m_{\frac{1}{2}}=160$\,GeV,
$A_0=100$\,GeV, $\tan\beta=40 $ and $\mu<0$; (2) $m_0=150$\,GeV,
$m_{\frac{1}{2}}=180$\,GeV, $A_0=300$\,GeV, $\tan\beta=40 $ and $
\mu>0$. \label{pt2}}
\end{figure}
\end{document}